\documentclass{aastex63}
\usepackage{overpic}

\received{June 1, 2019}
\revised{January 10, 2019}
\accepted{\today}
\submitjournal{ApJ}
\shorttitle{Differential rotation of the stellar halo}
\shortauthors{Tian et al.}
\graphicspath{{./}{figures/}}
\begin{document}
\title{Differential rotation of the halo traced by the K-giant stars}
\correspondingauthor{Chao Liu}
\email{liuchao@nao.cas.cn}
\author[0000-0003-3347-7596]{Hao Tian}
\affil{Key Laboratory of Space Astronomy and Technology, 
National Astronomical Observatories, 
Chinese Academy of Sciences, 
Beijing 100101, PR China;}
\collaboration{1}{(LAMOST Fellow)}

\author[0000-0002-1802-6917]{Chao Liu}
\affil{Key Laboratory of Space Astronomy and Technology, 
National Astronomical Observatories, 
Chinese Academy of Sciences, 
Beijing 100101, PR China;}
\nocollaboration{2}

\author[0000-0003-0631-568X]{Yougang Wang}
\affil{Key Laboratory of Computational Astrophysics, 
National Astronomical Observatories, Chinese Academy of Sciences, 
Beijing, 100101 PR China}
\nocollaboration{2}

\author[0000-0002-2459-3483]{Yan Xu}
\affil{Key Laboratory of Optical Astronomy, 
National Astronomical Observatories, 
Chinese Academy of Sciences, 
Beijing 100101, PR China;}
\nocollaboration{2}

\author[0000-0003-1972-0086]{Chengqun Yang}
\affil{Key Laboratory of Optical Astronomy, 
National Astronomical Observatories, 
Chinese Academy of Sciences, 
Beijing 100101, PR China;}
\nocollaboration{2}

\author[0000-0002-6434-7201]{Bo Zhang}
\affil{Department of Astronomy, Beijing Normal University, Beijing 100875, PR China}
\collaboration{1}{(LAMOST Fellow)}

\author[0000-0002-0642-5689]{Xiang-Xiang Xue}
\affil{Key Laboratory of Optical Astronomy, 
National Astronomical Observatories, 
Chinese Academy of Sciences, 
Beijing 100101, PR China;}
\nocollaboration{2}
\begin{abstract}

We use K-giant stars selected from the LAMOST DR5 to study the variation of the rotational velocity  
of the galactic halo at different space positions.
Modelling the rotational velocity distribution with both the halo and disk components, 
we find that the rotational velocity of the halo population decreases almost linearly 
with increasing vertical distance to the galactic disk plane, $Z$, at fixed 
galactocentric radius, $R$. The samples are separated into two parts with 
 $6<R<12$ kpc and $12<R<20$ kpc. We derive that  the 
decreasing rates along $Z$ for the two subsamples are $-3.07\pm0.63$ and
$-1.89\pm0.37$ km s$^{-1}$ kpc$^{-1}$, respectively.
Compared with the TNG simulations, we suggest that this
trend is probably caused by the interaction between
the disk and halo. The results from the simulations show that only the oblate halo 
can provide a decreasing rotational velocity with an increasing $Z$. This indicates that
the Galactic halo is oblate with galactocentric radius $R<20$ kpc.
On the other hand, the flaring of the disk component 
(mainly the thick disk) is clearly traced by this
study, with $R$ between 12 and 20 kpc, the disk can vertically extend 
to $6\sim10$ kpc above the disk plane.
What is more interesting is that, we find the 
 Gaia-Enceladus-Sausage (GES) component has a 
significant contribution only in the halo with $R<12$ kpc, i.e. a fraction of 23$-$47\%.
While in the outer subsample,
the contribution is too low to be well constrained.
\end{abstract}


\section{Introduction} 

The stellar halo is  one of the most important components in the Milky Way. 
Under the paradigm of $\Lambda$ cold dark matter
model,  the halo is formed through accretion and merging the satellites, and  plenty of substructures
are supposed to be left in the halo. So the halo has been recording 
the information of the formation history. As a result, studies on the stellar halo can directly help
us understand the formation our Milky Way. 
But the stellar halo is  the most difficult component 
to be studied. The halo is quite diffused and of low density, and
it can reach out to the volumes as far as more
than 100 kpc \citep[also references 
there]{2016ARA&A..54..529B}. That means 
it is hard to obtain the velocity information, i.g. proper 
motions and radial velocities. What is more,
the distance is also difficult to be well determined, except 
the standard candles like RR Lyrae stars or the blue 
horizontal branch stars \citep{Xue2008ApJ...684.1143X, 
Hernitschek2018ApJ...859...31H, Thomas2018MNRAS.481.5223T}. 
That brings the main difficulty to obtain
a sufficient sample of tracers
to study the properties of the stellar halo. 

Thanks to the fast development of large survey projects, e.g. the Sloan Digital Sky Survey
\cite[SDSS;][]{2000AJ....120.1579Y}, the Panoramic 
Survey Telescope and Rapid Response System
\cite[Pan-STARRS;][]{2016MNRAS.463.1759B} and 
Gaia mission \citep{2016A&A...595A...1G, 2018A&A...616A...1G}, many of 
those embedded substructures have been discovered  in the halo,
e.g. the Sagittarius Stream \citep{1994Natur.370..194I}, GD-1 Stream
\citep{2006ApJ...641L..37G, 2006ApJ...643L..17G} and 
the $\omega$-Cen Stream \citep{2019NatAs...3..667I}.  Those substructures, 
especially those thin cold streams,
are quite helpful for studying the halo profile. 
\citet{2013MNRAS.436.2386L} introduced a method using
Markov Chain Monte Carlo technique to constrain the 
halo shape with thin streams. The streams 
NGC 5466 and Pal 5 are proved to be the best candidate according to their orbit properties. 
\citet{2015ApJ...801...98S} also studied how to constrain the halo profile 
with action distributions of streams. The results show that even for the simple case of spherical potential,
at least 20 streams with more than 100 member stars are required to make sure the 
potential well constrained. 
With the most prominent stream, the Sagittarius Stream,
\citet{2010ApJ...714..229L} introduced a triaxial model, which can well reproduce the Sagittarius Stream. 
But there are still some points which are not consistent with observations in the following years 
after that \citep[see][and references mentioned there]{2017ApJ...836...92D}.  \citet{2013ApJ...773L...4V}
also studied the halo shape using the Sagittarius Stream with the effect of the
Magellanic Clouds considered. Differently, their results suggested an oblate halo. According to all above,
we can find that the streams are powerful tracers to constrain the halo profile. 

Many direct efforts other than using tidal substructures 
are also made to profile the halo. \cite{2012MNRAS.419.1951V} showed that the halo shape can be 
probed with the orbital properties of individual halo stars, e.g. the action and frequency. 
The results from the 
complementary simulations show that the disk plays an important role to the 
shape of the inner halo and making it oblate, but not for the outer part.  
Using the K-giant stars selected from the DR5 of the Guoshoujing Telescope
(Large Sky Area Multi-Object Fiber Spectroscopic Telescope, LAMOST hereafter)
, \citet{2018MNRAS.473.1244X} showed a complicated halo, the profile is different for 
the inner and outer parts. Traced by the K-giant 
stars selected from the LAMOST DR5, the shape of the halo varies from oblate for the inner part to almost 
spherical for the outer part.

To figure out the interaction between different components in the Milky Way, 
we should do a deep analysis on the dynamics of the components. 
The second data release of Gaia mission (Gaia DR2, hereafter) contains 
proper motions and parallaxes for more than 
1.3 billion stars, and radial velocities for stars brighter than 14
in \emph{G}-band \citep{2018A&A...616A...1G}.
The accurate astrometric data greatly improves the study on the 
dynamics of the disk and the halo\citep{Belokurov2019arXiv190904679B}. 
The phase spiral signature in the local volume was discovered
by \cite{2018A&A...616A..11G} and \citet{2018Natur.561..360A} for the first time, 
which indicates a possible interaction between
the satellite of the Milky Way and the disk \citep[Xu et al. in prep.]{2019MNRAS.485.3134L}.  
The spectroscopic surveys, including APOGEE and LAMOST, 
have brought a big progress during the study on the 
Milky Way.  The combination of the APOGEE/LAMOST and the Gaia datasets
provides a unique opportunity 
to study the formation of the halo. 
Combining  the astrometric data Gaia DR2 and APOGEE, 
\cite{2018Natur.563...85H} revealed a major merger event 
in the local volume, named Gaia-Enceladus
\citep[also known as Gaia-Sausage,][]{Belokurov2018MNRAS.478..611B}. 
With the combination of  spectroscopic survey LAMOST and Gaia DR2,
\citet[hereafter Paper I]{2019ApJ...871..184T}   
measured the halo rotational velocity  of $V_T=+27^{+4}_{-5}$ km s$^{-1}$ of the halo in
the solar neighbourhood  using the K-giant sample. 

According to the studies by \citet{2017MNRAS.467.3083R}, the morphology
for the massive and dwarf galaxies  significantly depends
on the assemble history and spin, 
respectively. For the Milky Way-like galaxies, the morphology 
depends on a combination of the two 
factors. To figure out if the halo shape is related to the spin variance, 
similar with Paper I, we will also use the K-giant stars to do a 
further study on the rotation of the halo.
The K-giant stars are perfect tracers for studying the dynamics of the halo. First,
giant stars have higher luminosity, which is quite helpful to trace distant volumes.
Second, the K-giant stars are of higher quantity which brings enough samples for 
statistics, especially for studies on the global properties of the stellar halo. 

The LAMOST is a 4 meter, quasi-meridian, 
reflecting Schmidt telescope. There are
4000 fibers, which make it  efficient to obtain spectra. 
 The fifth data release (DR5) 
provides the radial velocity and metallicity for millions
of stars with uncertainties of $\sim$5 km s$^{-1}$ and 0.1 dex 
respectively. The high efficiency 
helps LAMOST obtain more than 9 million spectra, which is the largest 
spectrum observation sample. Combined with Gaia DR2, it provides
the unprecedented chance to study the Milky Way. More recently, the Seventh Data Release
includes low- and medium-resolution spectra\footnote{http://dr7.lamost.org/} 
of $R=1800$ and 7500, respectively.

This paper is constructed as follows.
In Section \ref{Sec:DM}, we briefly introduce the dataset and the 
method. The results are shown in Section \ref{Sec:results}. 
The discussions are listed in Section
\ref{Sec:Disc}.

\section{Data \& Method} \label{Sec:DM}
The same with Paper I, we  still use the K-giant stars
selected from LAMOST DR5 \citep{2014ApJ...790..110L} to study the rotation velocities of the halo and thick 
disk in this paper. The LAMOST DR5 dataset
provides the radial velocity and metallicity, with typical errors 
of $\sim 5$ km s$^{-1}$ and 0.1 index, respectively. Following Paper I,
we remove all the stars with metallicity
[Fe/H]$>-1$ to reduce the contamination
from disk, especially the thin disk stars \citep[See Figure 4 in][]{2015ApJ...808..132H}.  
That is proved to be quite efficient (as shown in Paper I). In this project, 
we aim at larger volumes, where the distances of the distant stars are
no longer available from Gaia DR2. We adopt the distances provided by \cite{2015AJ....150....4C},
which are estimated using the 
technique of Bayesian approach by comparing the stellar parameters and a grid of stellar isochrones
with typical relative errors  $\sim$ 20\%.

To avoid systematic offset of the distances and radial velocities 
from LAMOST DR5 \citep[][and Paper I]{2015ApJ...809..145T, 2017MNRAS.472.3979S},
we use the common stars of both LAMOST DR5 and Gaia DR2 to 
figure out the offset and correct the values from LAMOST DR5. The distance is  normalized 
by  0.805 and the radial velocity is corrected by adding $\sim$5.38 km s$^{-1}$. 
More details for distance and radial velocity correction are described in 
Appendix~\ref{sec:D_correct} and \ref{sec:RV_correct}, respectively.

\begin{figure}
 \centering
 \includegraphics[width=0.46\textwidth]{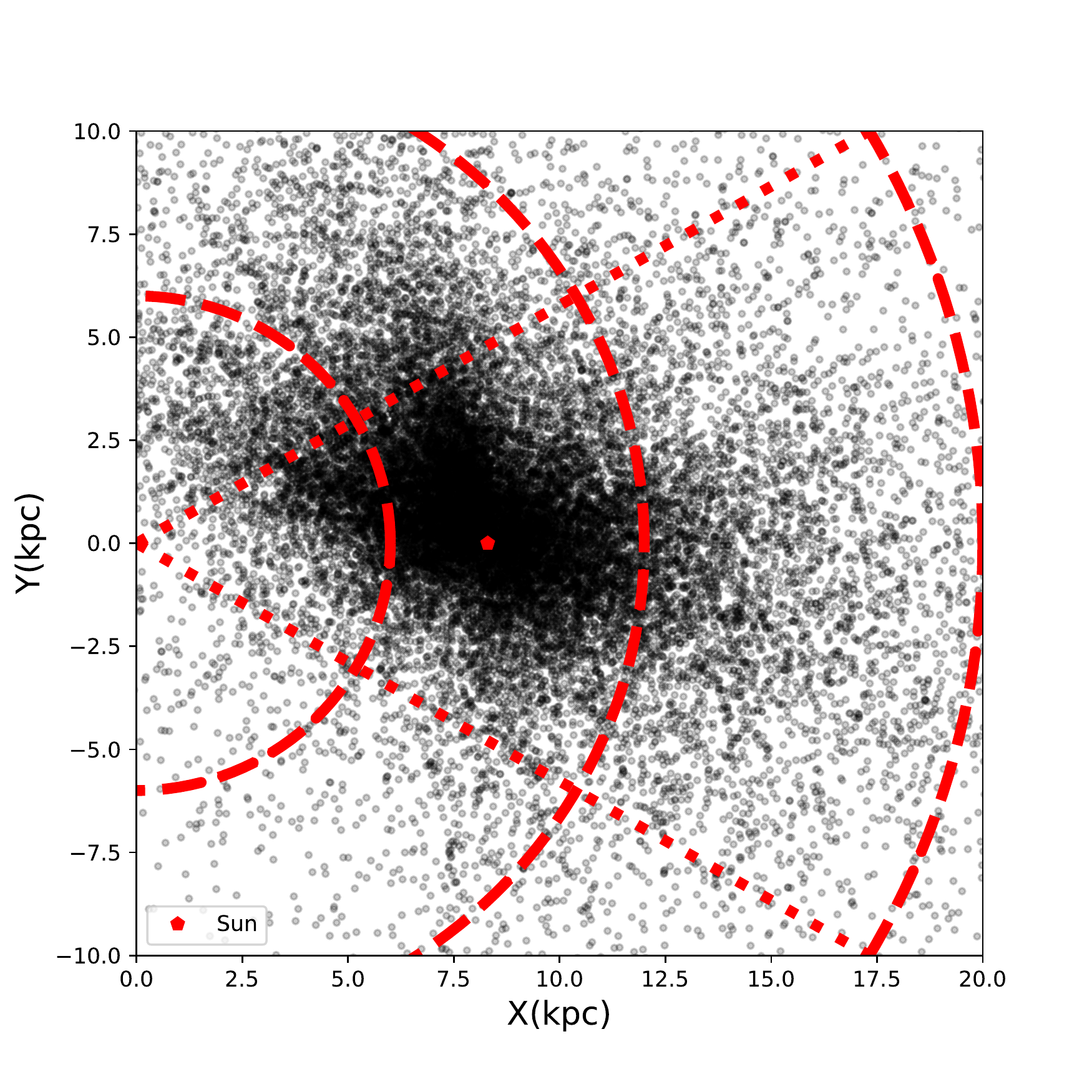}
 \caption{The space distribution of the K-giant sample is shown. 
 The dashed and dotted lines represent the limits
 on $\theta$ and $R$ in cylindrical frame, $-30^\circ$ and $30^\circ$ for $\theta$ and 6, 12 and 20 kpc for R from inner to outer.}
 \label{Fig:space}
\end{figure}

\begin{table*}
\footnotesize
    \centering
    \begin{tabular}{c|c|c|c|c|c|c|c|c|c|c|c}
    \hline \hline
    & Selection            & $f_{H}$             & $V_\phi^{H}$        & $\sigma_{V_\phi}^{H}$        &$f_{D}$              & $V_\phi^{D}$       & $\sigma_{V_\phi}^{D}$     & $f_{H^3}$   & $V_\phi^{H^3}$         & $\sigma_{V_\phi}^{H^3}$      &   $N$ \\ 
    &                             &                   & km s$^{-1}$  & km s$^{-1}$  &                   & km s$^{-1}$  & km s$^{-1}$  &      & km s$^{-1}$    & km s$^{-1}$   &       \\ \hline 
  S&  $-1<Z<1$    & $0.70^{+0.04}_{-0.04}$  & $37^{+7}_{-7}$      & $75^{+4}_{-4}$     &$0.30$           & $185^{+5}_{-5}$   & $ 41^{+3}_{-3}$   & -         & -                 & -            &    2215 \\
    &~$1<Z<2$     & $0.81^{+0.04}_{-0.05}$  & $ 50^{+5}_{-6}$      & $79^{+3}_{-4}$     &$0.19$           & $177^{+5}_{-6}$   & $38^{+4}_{-4}$   & -         & -                 & -            &    2268 \\
    &~$2<Z<4$     & $0.12$  &$-18^{+41}_{-64}$      &  $82^{+17}_{-16}$    &$0.41^{+0.06}_{-0.07}$  & $127^{+9}_{-9}$    & $55^{+4}_{-4}$   & $0.47^{+0.09}_{-0.11}$  &  $12^{ +4}_{-4}$    & $33^{+5}_{-7}$   &    3415 \\
    &~$4<Z<6$     & $0.77^{+0.05}_{-0.09}$  & $40^{+6}_{-4}$      & $73^{+3}_{-3}$     & -                 & -              & -           & $0.23                $  & $6^{+7}_{-7}$          & $18^{+16}_{-12}$   &    1731 \\
    &~$6<Z<10$   & $0.68^{+0.05}_{-0.05}$  & $24^{+4}_{-4}$      & $82^{+4}_{-3}$     &-                  & -              & -           & $0.32                $  & $14^{+4}_{-5}$       & $12^{+10}_{-8}$   &    1327 \\ 
    &~$10<Z<15$ & 1                 & $10^{+4}_{-4}$      & $63^{+3}_{-3}$    &-                  & -               & -           &  -        & -               & -             &     496 \\ \hline \hline
  SO&  $-1<Z<1$    & $0.60^{+0.04}_{-0.04}$  & $23^{+11}_{-10}$     & $80^{+13}_{-12}$   &$0.40$          & $221^{+2}_{-3}$     & $ 20^{+3}_{-3}$   &  -        & -               & -             &    287 \\
    &~$1<Z<2$     & $0.86^{+0.03}_{-0.04}$  & $ 20^{+7}_{-6}$       & $59^{+10}_{-7}$   &$0.14$          & $201^{+7}_{-10}$   & $28^{+11}_{-8}$  &  -        & -               & -             &    218 \\
    &~$2<Z<4 $     & $0.92^{+0.02}_{-0.02}$  & $20^{ +3}_{-3}$       & $46^{+4}_{-4}$     &$0.08$          & $194^{+8}_{-14}$   & $32^{+11}_{-6}$  &  -        & -               & -            &    537 \\
    &~$4<Z<6$     & $0.95^{+0.01}_{-0.02}$  & $9^{ +4}_{-4}$         & $54^{+4}_{-4}$     &$0.05$          & $201^{+9}_{-17}$   & $28^{+16}_{-8}$  & -         & -               & -             &    482 \\
    &~$6<Z<10$   & $0.97^{+0.01}_{-0.01}$  & $6^{+3}_{-3}$         & $53^{+3}_{-3}$     &$0.03$          & $192^{+18}_{-26}$   & $45^{+22}_{-17}$  & -         & -               & -             &    639 \\ 
    &~$10<Z<15$ & 1                & $1^{+4}_{-4}$        & $60^{+4}_{-4}$      & -                 & -               & -           & -         & -               & -             &    395 \\ \hline \hline
    \end{tabular}
    \caption{The first column lists the selection for each subsample. 
    The results, including fraction $f$, the rotational velocity $<V_\phi>$ and dispersions
    $\sigma_{V_\phi}$, from Bayesian Method for each component are listed. Symbols \emph{H}, \emph{D} and 3
    denote the disk, halo and the third component, respectively. The last column lists the 
    numbers of the K-giant stars in each sub-sample. The top and bottom parts list
    the information of the volumes in S-sample and SO-sample, respectively.  }
    \label{tab:MCMC_results}
\end{table*}

\begin{figure*}
 \centering
 \includegraphics[width=0.46\textwidth]{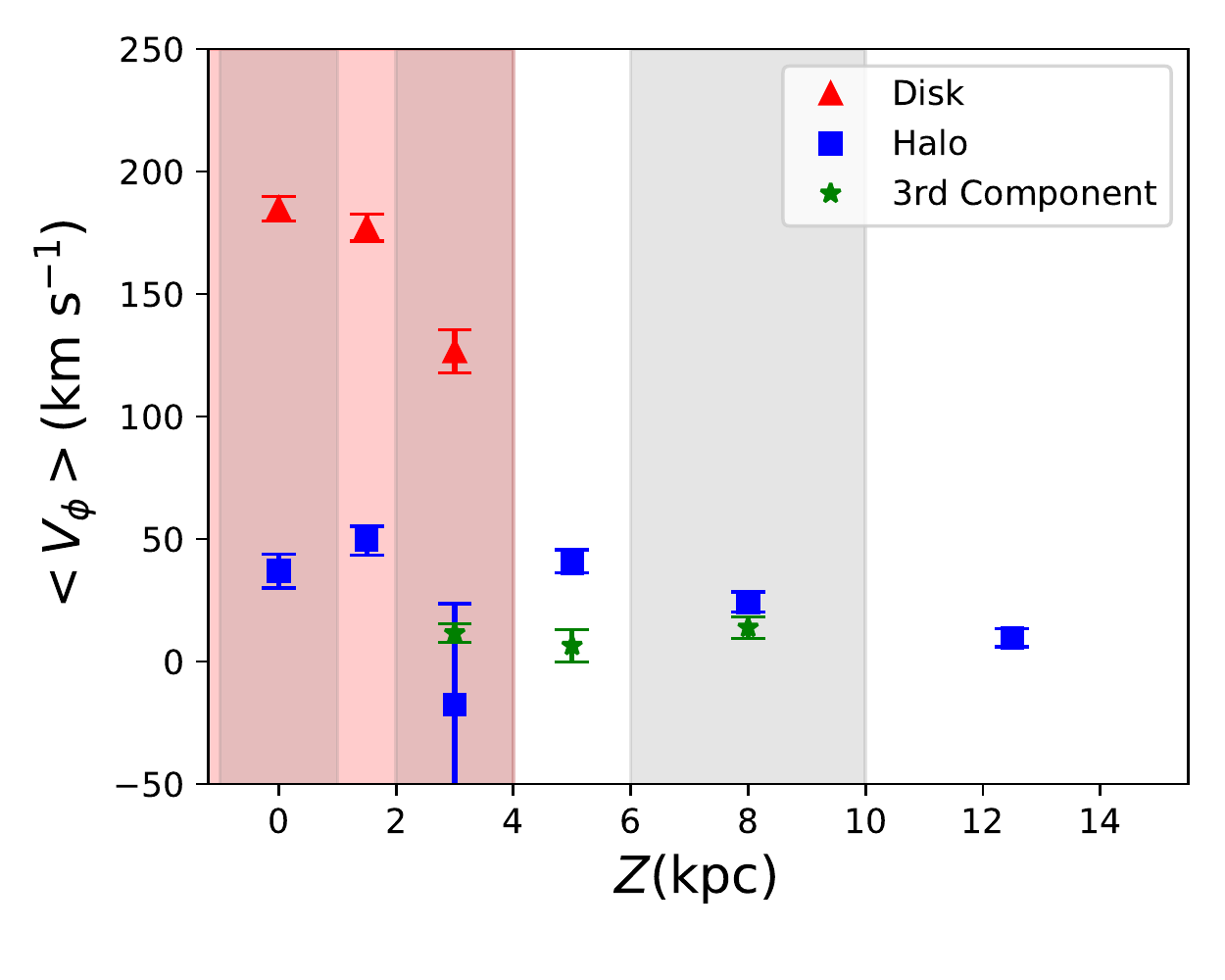} 
 \includegraphics[width=0.46\textwidth]{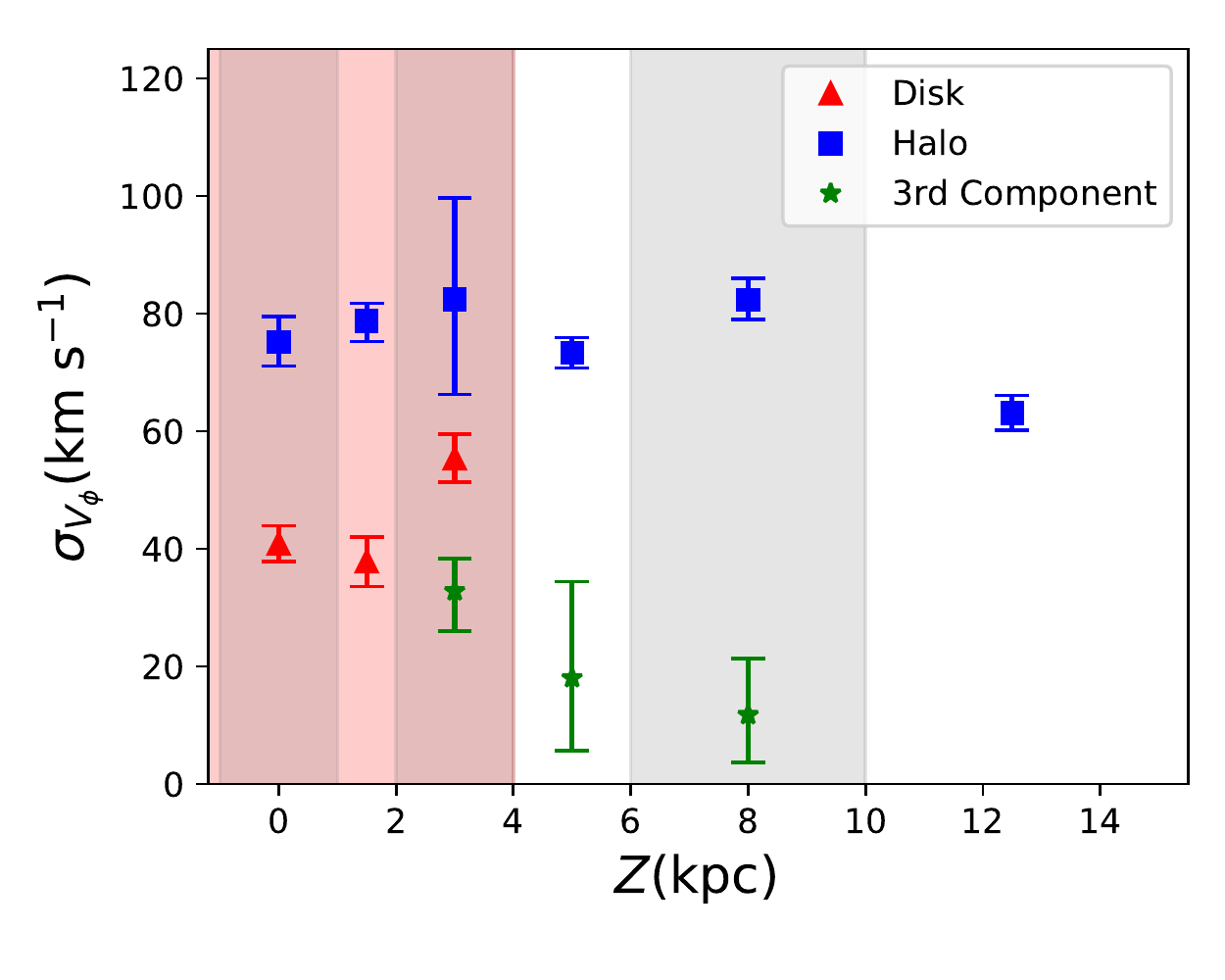}\\ 
 \vspace{-1.25cm}
 \includegraphics[width=0.46\textwidth]{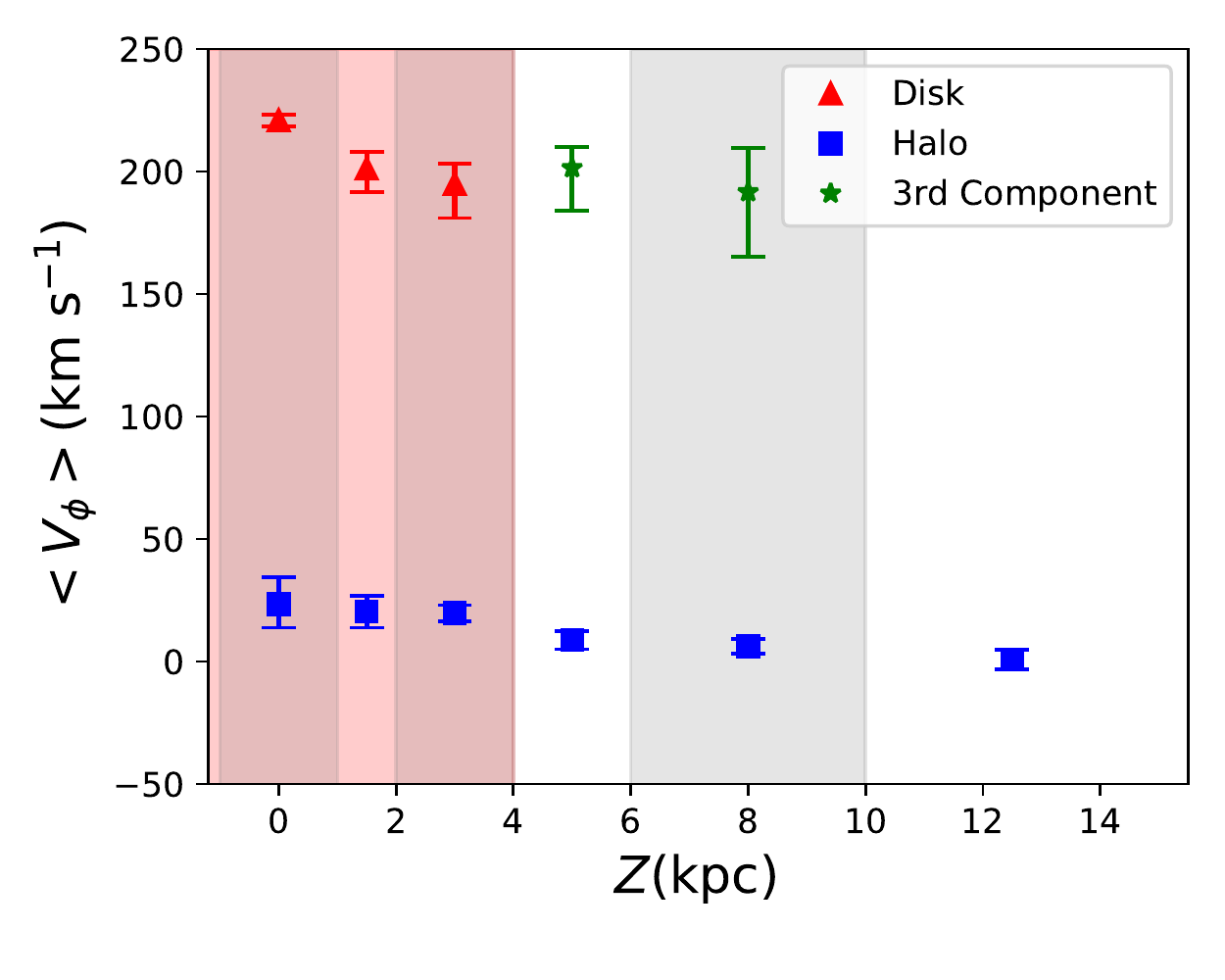}
 \includegraphics[width=0.46\textwidth]{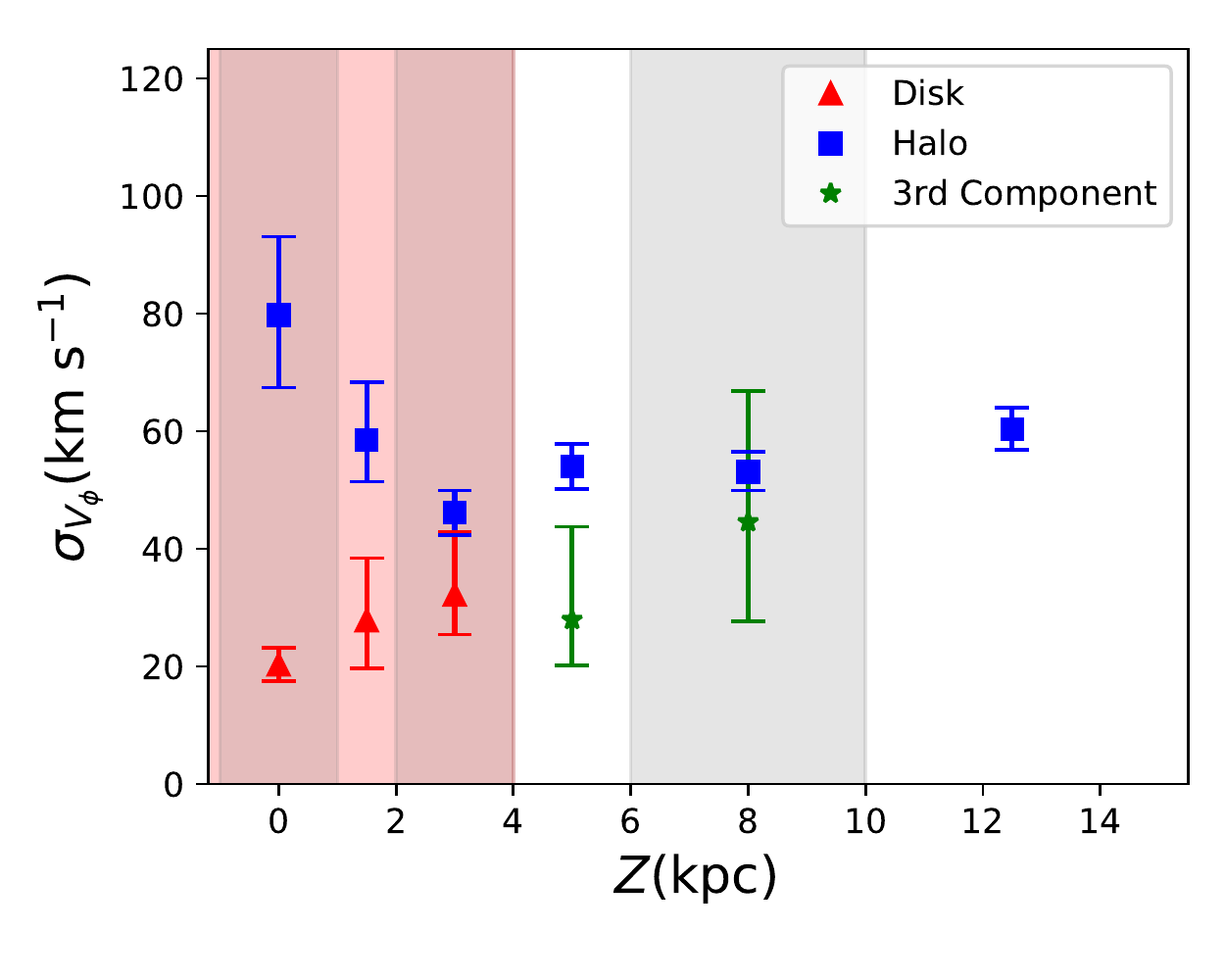}\\ 
 \caption{The distributions of the rotation velocity and its dispersion are shown in the left and right panels, respectively.
  The gray shadowed columns represent the bin range along $Z$. 
 The red shadowed region represent the height lower than 4 kpc,
 where the disk component is always included in the Bayesian model.
 The top and bottom panels show the results from the S-sample and SO-sample, respectively. The blue 
 and red symbols represent the parameters of the halo and disk components, respectively. The green ones represent the 
 third component in the model. The errorbars represent the uncertainties
 of the values which are obtained from Markov chain Monte Carlo simulations.}
 \label{Fig:variance}
\end{figure*}

Proper motions of the K giant star samples are obtained by cross-matching with Gaia DR2, then 
the positions and velocities are calculated with 
Python package \emph{Galpy} \citep{2015ApJS..216...29B}. 
The Solar motion relative to the Local Standard of Rest from \cite{2010MNRAS.403.1829S} is adopted,
$(U_\odot, \,V_\odot,\, W_\odot)=(11.1,\,12.24, \,7.25)$ km s$^{-1}$.

In order to study the variance of the rotational velocity of the halo
with different heights to the disk plane, a sufficient sample for each volume is required. In this
way, we firstly focus
on the volumes with $6<R<12$ kpc, $-30^{\circ}<\phi<30^{\circ}$, and $-1<Z<15$ kpc in galactocentric centered 
cylindrical coordinates.  We adopt the Solar location from \cite{2014ApJ...783..130R} with ($R$, $\phi$, $Z$)=(8.3 kpc, 0$^{\circ}$, 0 kpc). 
 Here we ignore the distance of the Sun to the disk plane, which is too small to make any difference to our results.
To constrain the uncertainties, only those stars with signal to noise ratio higher than 10,
radial velocity errors
$\sigma_{RV}<10$  km s$^{-1}$ and proper motion errors
($\sigma_{\mu_\alpha}$, $\sigma_{\mu_\delta}$) smaller than
0.3 mas yr$^{-1}$ are used. This is labeled as the S-sample. A similar sample with different $R$ range,
$12<R<20$ kpc, 
is also selected for studying the outer part, labeled as SO-sample.

Along the height to the disk plane, we divide  each of the two
samples into 6 sub-volumes to make sure each
sub-volume contains enough number of stars. 
The information for each sub-volume is listed in Table \ref{tab:MCMC_results}.
The numbers of stars in each sub-volume are listed in the last column. 

In Paper I, a Bayesian  model including
three Gaussian components was used  for the local volume,
the halo, the thick disk and a possible 
retrogradely rotating component. In this paper, we   adopt the same  method that
includes the halo, the thick disk and a possible additional component. 
The Gaussian distribution of each component can be written as follow:
\begin{equation}\label{equ:gauss}
p_i(V_\phi^{(k)}|f_i,V_{\phi,i},\sigma_i,\epsilon^{(k)})={\frac{f_i}{\sqrt{2\pi(\sigma_i^2+\epsilon_k^2)}}}\exp\left(-\frac{(V_\phi^k-V_{\phi,i})^2}{2(\sigma_i^2+\epsilon_k^2)}\right),	
\end{equation}
 where $V_{\phi,i}$ and $\sigma_i$ are the rotational velocity 
and its dispersion of the $i$th component, $f_i$ is the number fraction
of the $i$th component. $V_\phi^k$ and $\epsilon_k$ are the 
rotational velocity and its uncertainty of the $k$th star in cylindrical coordinates. 
Considering that the errors of
the rotational velocity will be larger for distant stars,
which may affect results,  here we  use the probability calculation 
with uncertainties involved, as shown in Equation \ref{equ:gauss}.
To determine the rotational velocities of the three components, the Python package \emph{emcee} 
is applied on samples with different space selections as listed in Table~\ref{tab:MCMC_results}.  Affline solver is used with 50 walkers and 6000 iterations in total, including
3000 burn-in iterations. 
Trying to find the best-fit parameters with the components,
we test models with different numbers of components, 
e.g. 1, 2 or 3 components according to the results mentioned 
in Paper I. The median values are adopted for each of the parameters. The lower and upper
uncertainties are determined with the differences between the median value and the 16\% and 84\%
values. A typical result for the parameter determination is showed in the Section~\ref{Sec:emcee}.
 Due to the large contribution from the disk, the third component has
low fraction in the lower volumes,  especially in this paper, we have much 
lower number of stars in each volume than that in Paper I.

We find that 
 the model with the disk and the halo components is better for volumes with $Z<2$ kpc. 
 As the contamination of the disk becomes lower for higher volume,  i.e. $Z>4$ kpc, 
 the model with the halo and the retrogradely rotating components is better, rather than the halo and the disk. 
  In general,  the disk  contributes very few stars ($\sim 8\%$) in  those higher
$Z$ volumes ($Z=5$ kpc) around $R=9$ kpc  \citep{2018MNRAS.478.3367W}.
  As a result, 
  the contributions of the halo and  the additional contribution will rise. 
  In Table \ref{tab:MCMC_results}, for the transition volume  with $2<Z<4$ kpc in S-sample,
the disk contribution becomes weaker and the third component
( the GES as described in the following section) increases.
 Then we constrain the 
 parameters of a model with all the three components. This is not done for the volume 
 with similar height in SO-sample because of the low number of the samples
 and the low contribution of the  GES component (more discussion in the following sections).
 The median values are 
adopted as the best-fit parameters for each of the volume. 
The results for those 
space volumes are listed in  Table~\ref{tab:MCMC_results}. The markers $H$, $D$ and 3 denote
the halo, the disk and the third component, respectively.

\section{Results}\label{Sec:results}
Figure~\ref{Fig:variance} represents the distributions of the rotational velocity and 
its dispersion in the left and right panels, respectively. The top and bottom panels
are the results for  S-sample and SO-sample, respectively. In each panel,
the parameters for the disk and halo components are represented by the red and blue symbols, respectively.
The green ones show the parameters of the  GES component in S-sample, or the 
extension of the disk in SO-sample. 

\subsection{Rotational velocity distribution of the halo}

\begin{figure}
\begin{center}
\hspace{0.01cm}
    \includegraphics[width=0.8\textwidth]{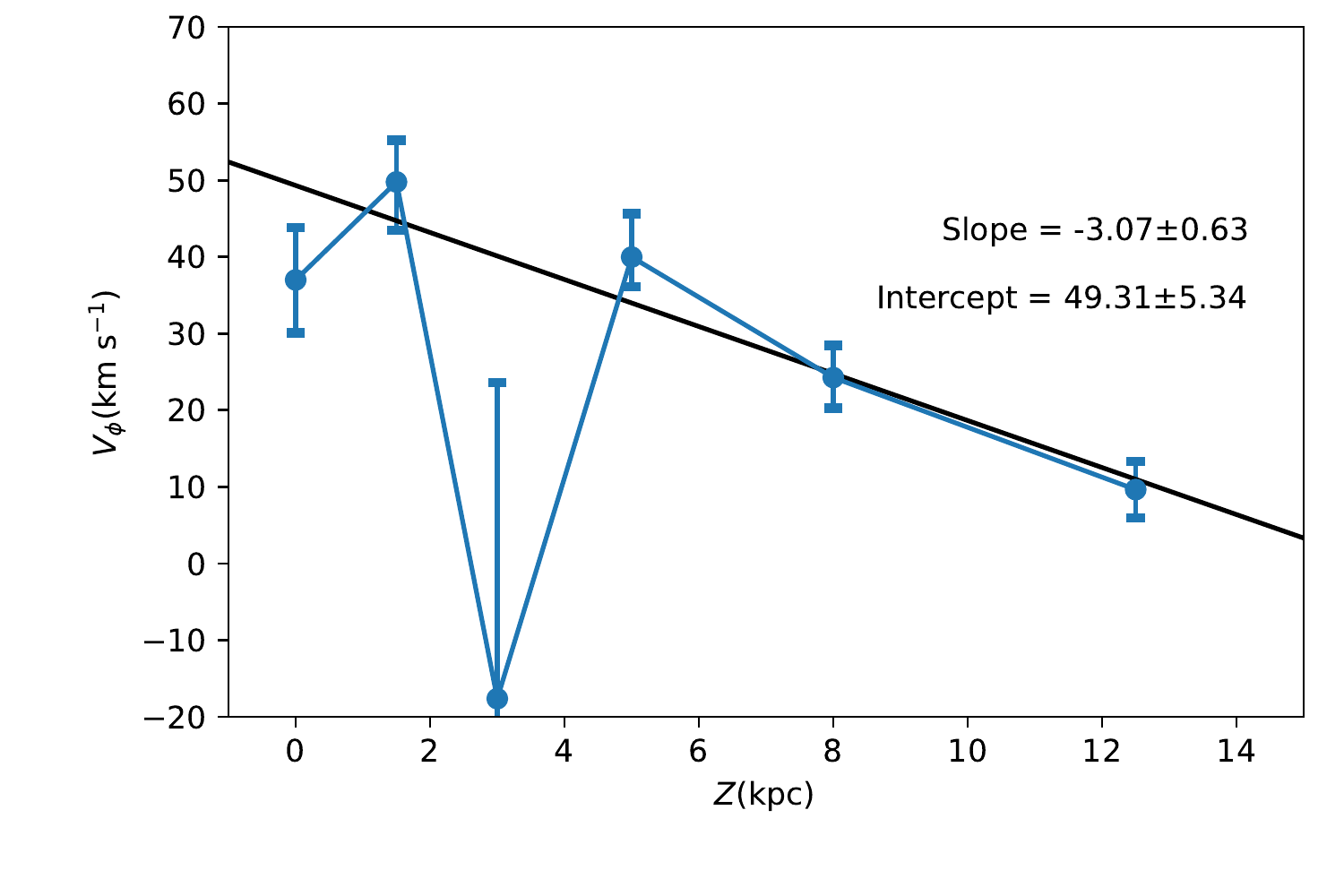}  \\ \vspace{-1.8cm}
    \includegraphics[width=0.8\textwidth]{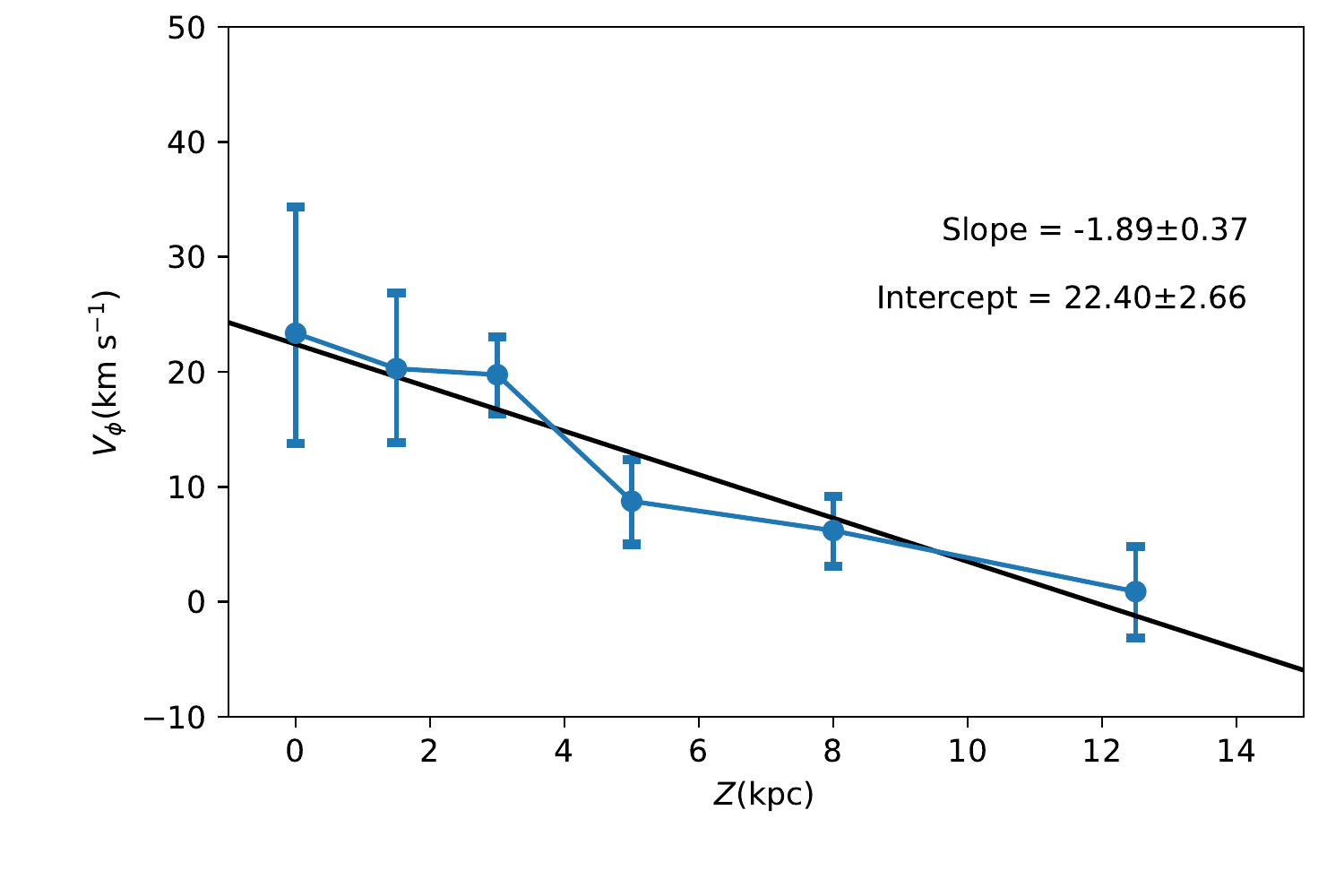}\\
    \caption{Rotational velocity distributions of the halo in the S-sample 
     and SO-sample are shown versus the height to the disk 
    in the left and right panels with the blue symbols, respectively. The solid black lines represent
    the regression linear relative. The parameters for the regression are marked on the top-right corner.}
    \label{fig:VT_Z_fit}
\end{center}
\end{figure}

Focusing on the halo component, which is the main target in this paper, 
we find that the halo in S-sample (the blue symbols in Figure~\ref{Fig:variance}) 
is progradely rotating with rotational velocity
dispersion around 75 km s$^{-1}$. The rotational velocity generally decreases 
versus the height to the disk plane. This is more clear in the left panel in Figure~\ref{fig:VT_Z_fit}.
The black solid line shows the linear fitting results, with rotational velocity uncertainties considered.
The fitting results show that the decreasing rate is  -3.04$\pm$0.63 km s$^{-1}$ kpc$^{-1}$. The intercept of the line
represents the rotational velocity of the halo at the disk plane, of 49$\pm$5.34 km s$^{-1}$. 
The volume with  $2<Z<4$ kpc shows an exception of the variance. 
 The uncertainties 
of the parameters are very large, that is because the fraction ($\sim 12\%$) 
of the halo is too low to be well constrained. 

Different with the S-sample, the rotational velocity dispersion is no longer flat  in the SO-sample, 
but decreases firstly and then keeps around a lower value, of 60 km s$^{-1}$. 
 This is consistent with the variance of the rotational velocity dispersion  
showed by \cite{2019AJ....157..104B}.

Similar to the S-sample,
 the SO-sample also shows a decreasing trend of the rotational velocity of the halo. As showed in the 
 right panel of Figure~\ref{fig:VT_Z_fit}, the decreasing rate of the rotational velocity is lower than
 that of S-sample, of -1.89$\pm$0.37 km s$^{-1}$ kpc$^{-1}$. The intercept of the fitting line is of 22.40$\pm$2.66 km s$^{-1}$. 
 This is much lower than that of S-sample. That means the rotational velocity of the halo 
close to the disk plane is larger for the inner part. 

\subsection{The contribution of GES}

The locations with $Z>2$ kpc show a second halo component with low rotational velocity dispersion 
and close-to-zero rotational velocity (the green symbols in the top panels). 
This is consistent with the properties of the GES
\citep{2018Natur.563...85H,  Belokurov2018MNRAS.478..611B, 2019MNRAS.488.1235M}. 
It should be noticed that the GES component is not recognized in the lower volumes, that does not mean
there is no contribution of the GES component in those volumes. The main reason is that the fraction of the 
GES member stars is too low to be well constrained  at lower $Z$ locations. 

\begin{figure}
\hspace{0.01cm}
    \includegraphics[width=0.42\textwidth]{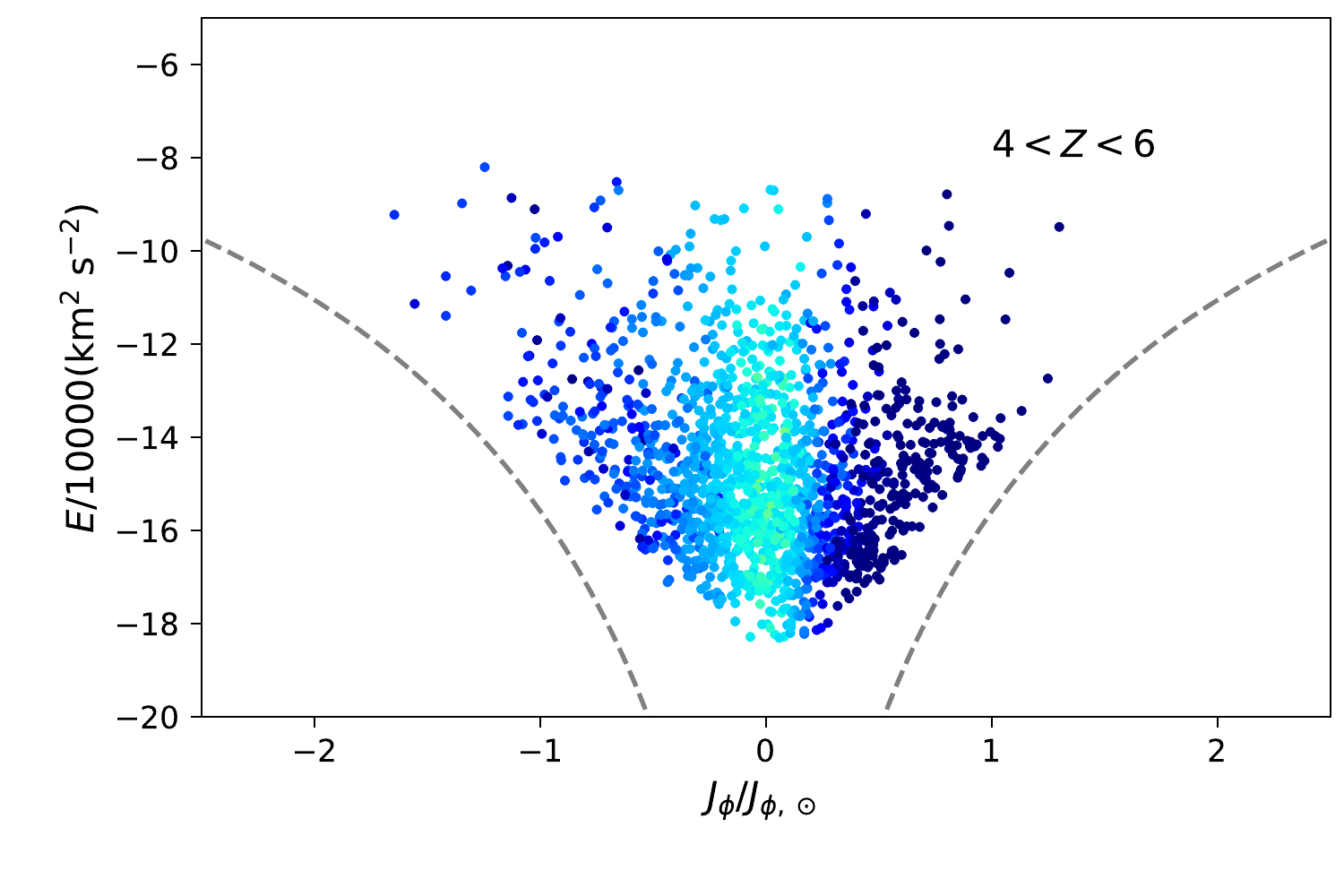} 
    \includegraphics[width=0.42\textwidth]{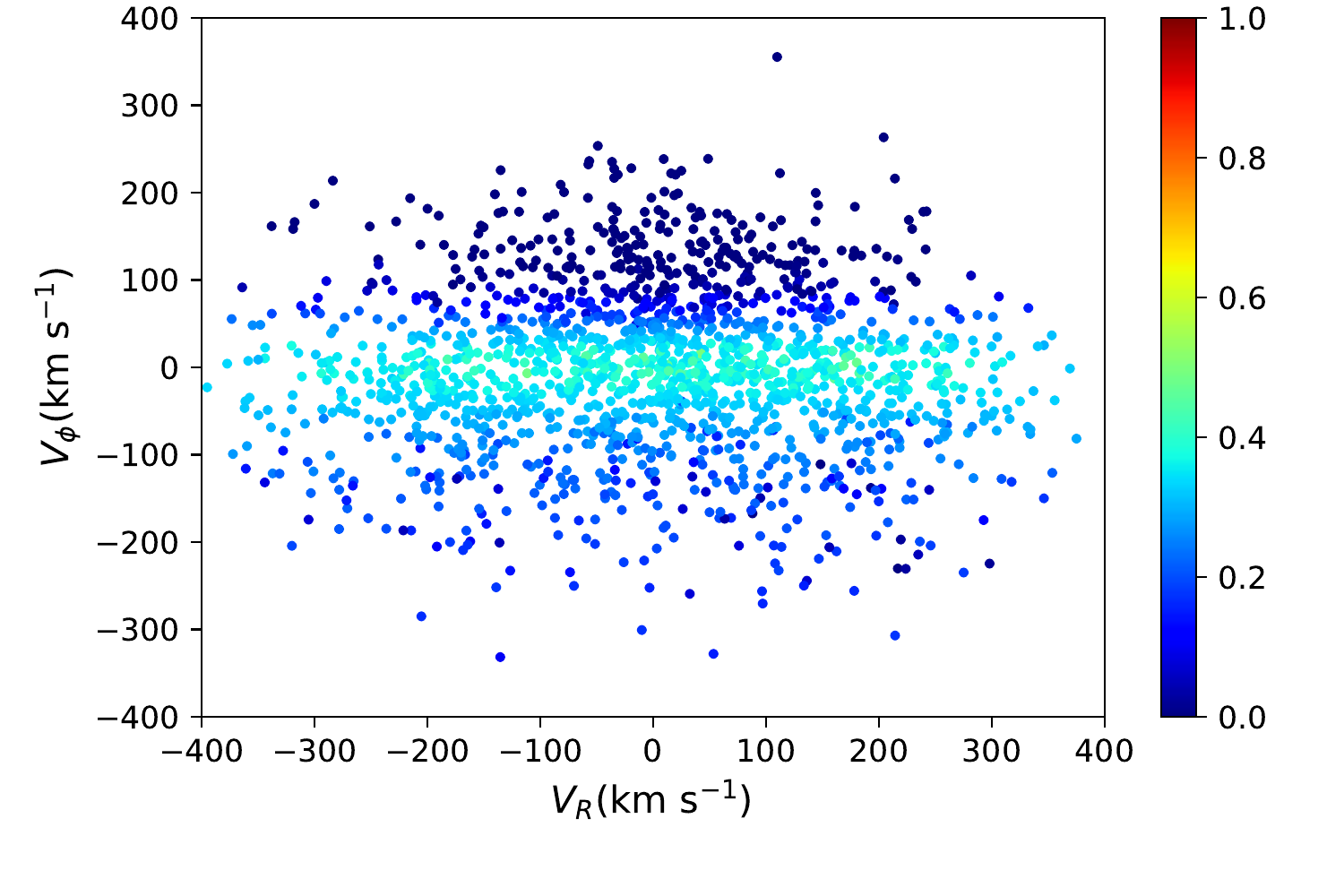}\vspace{-1.0cm}\\\vspace{-1.0cm}
    \includegraphics[width=0.42\textwidth]{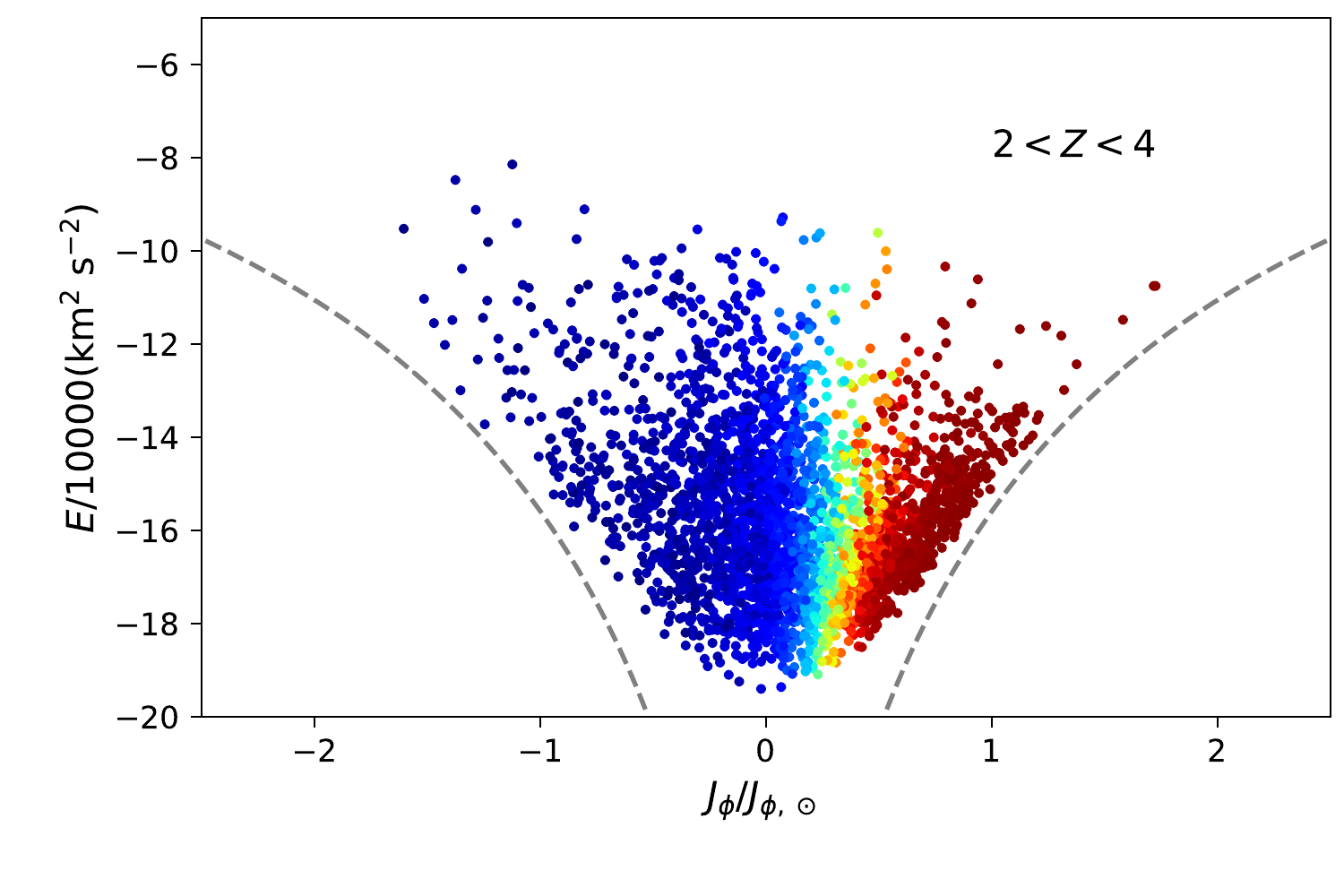} 
    \includegraphics[width=0.42\textwidth]{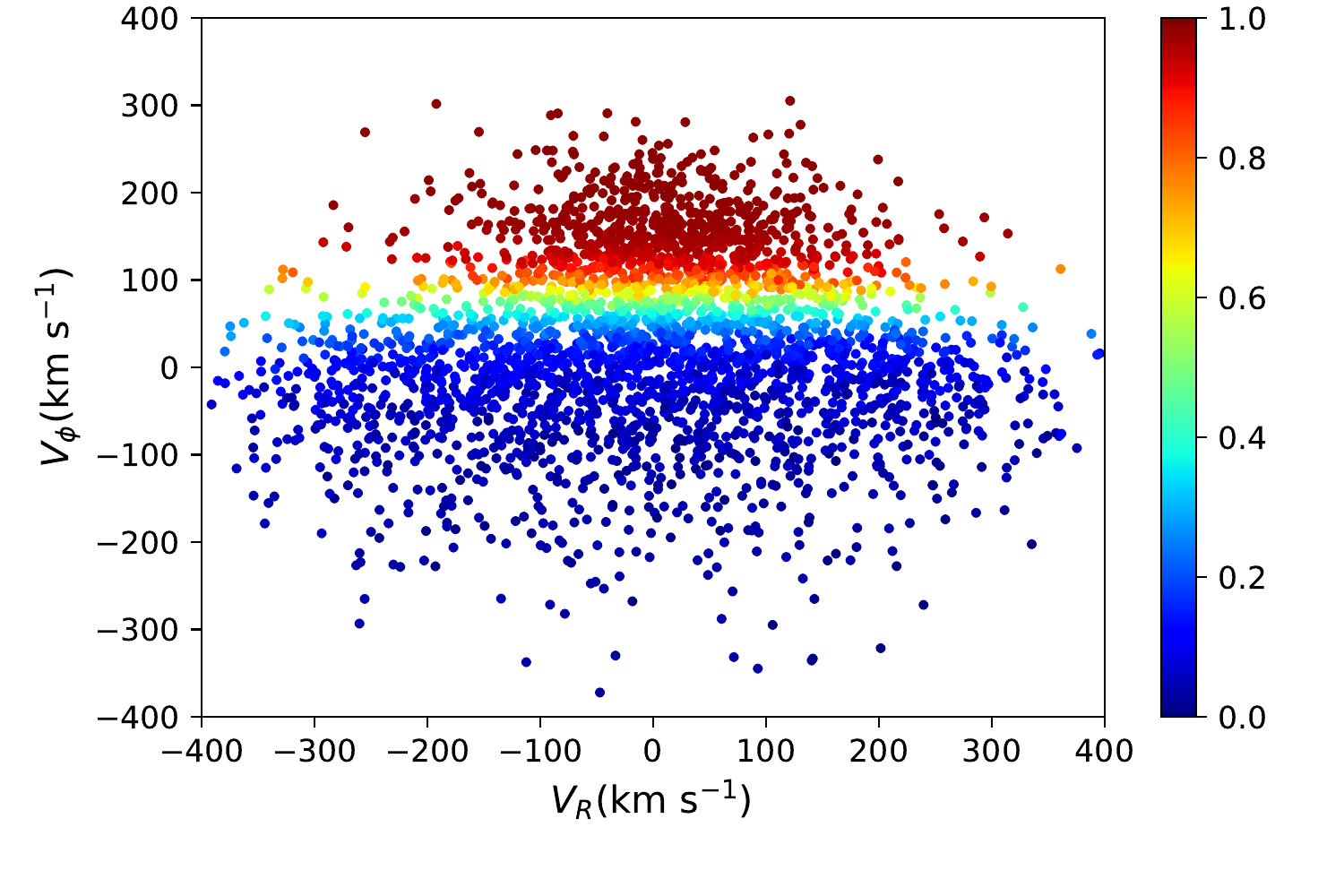}\\\vspace{-1.0cm}
    \includegraphics[width=0.42\textwidth]{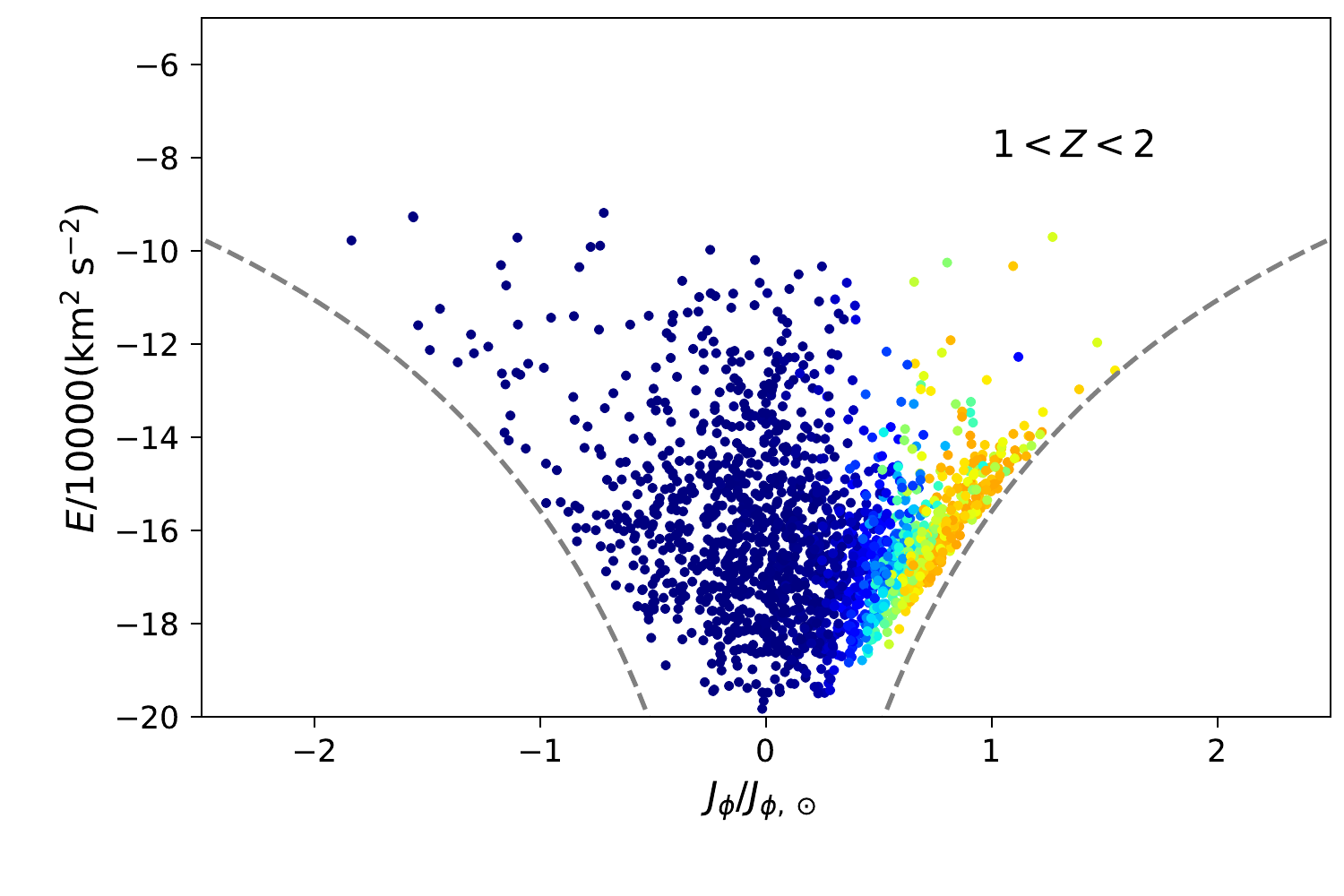}
    \includegraphics[width=0.42\textwidth]{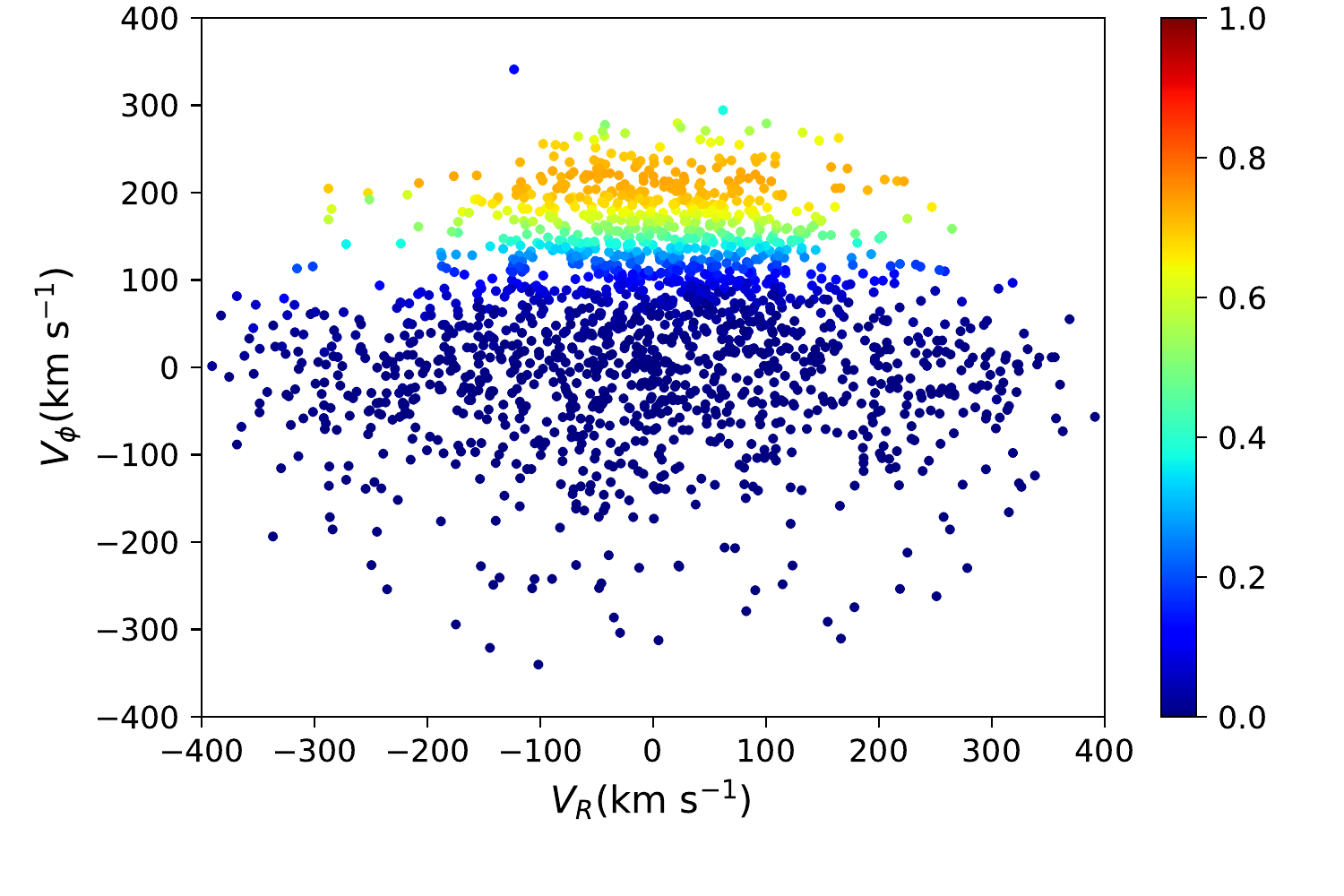}\\\vspace{-1.0cm}
    \includegraphics[width=0.42\textwidth]{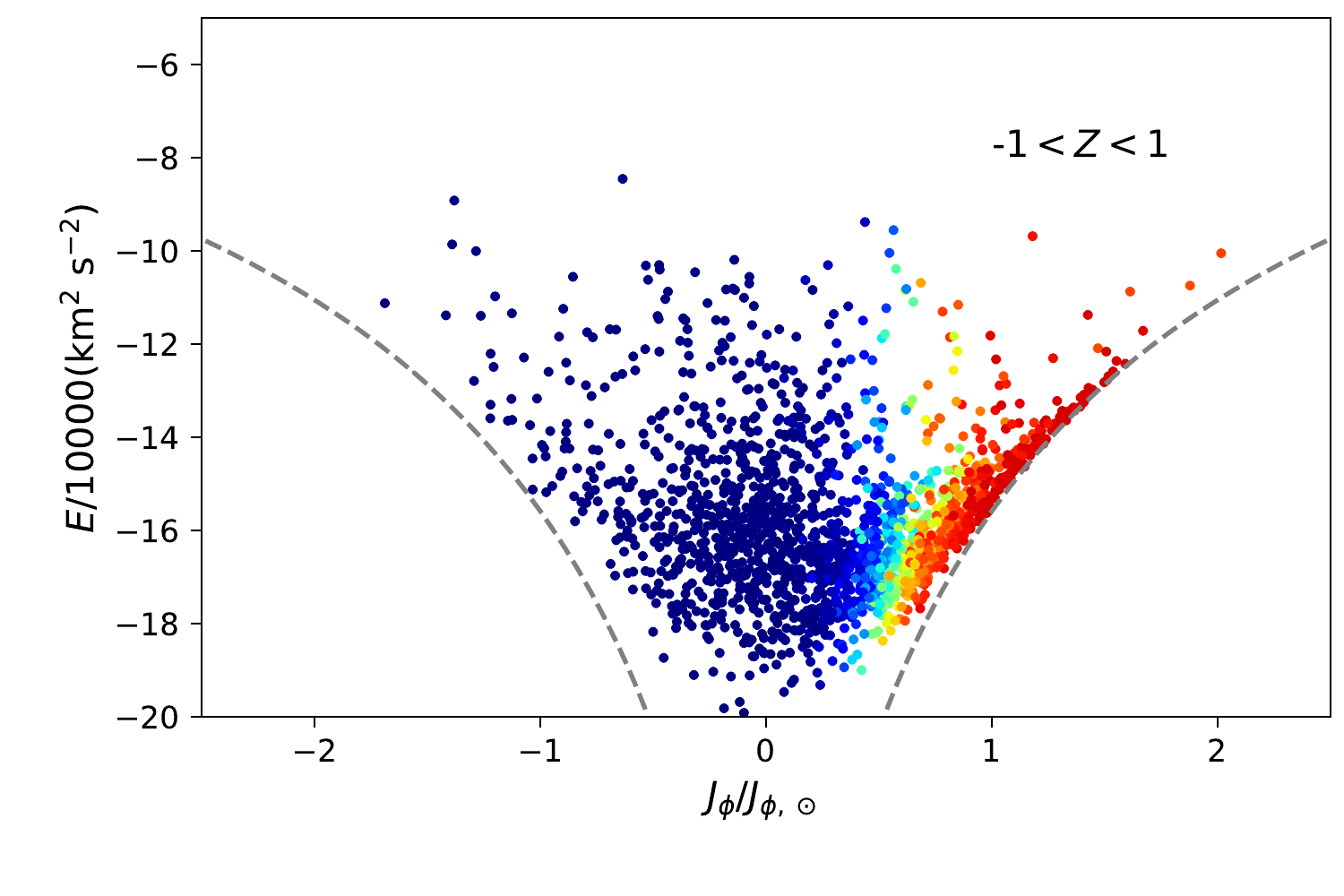} 
    \includegraphics[width=0.42\textwidth]{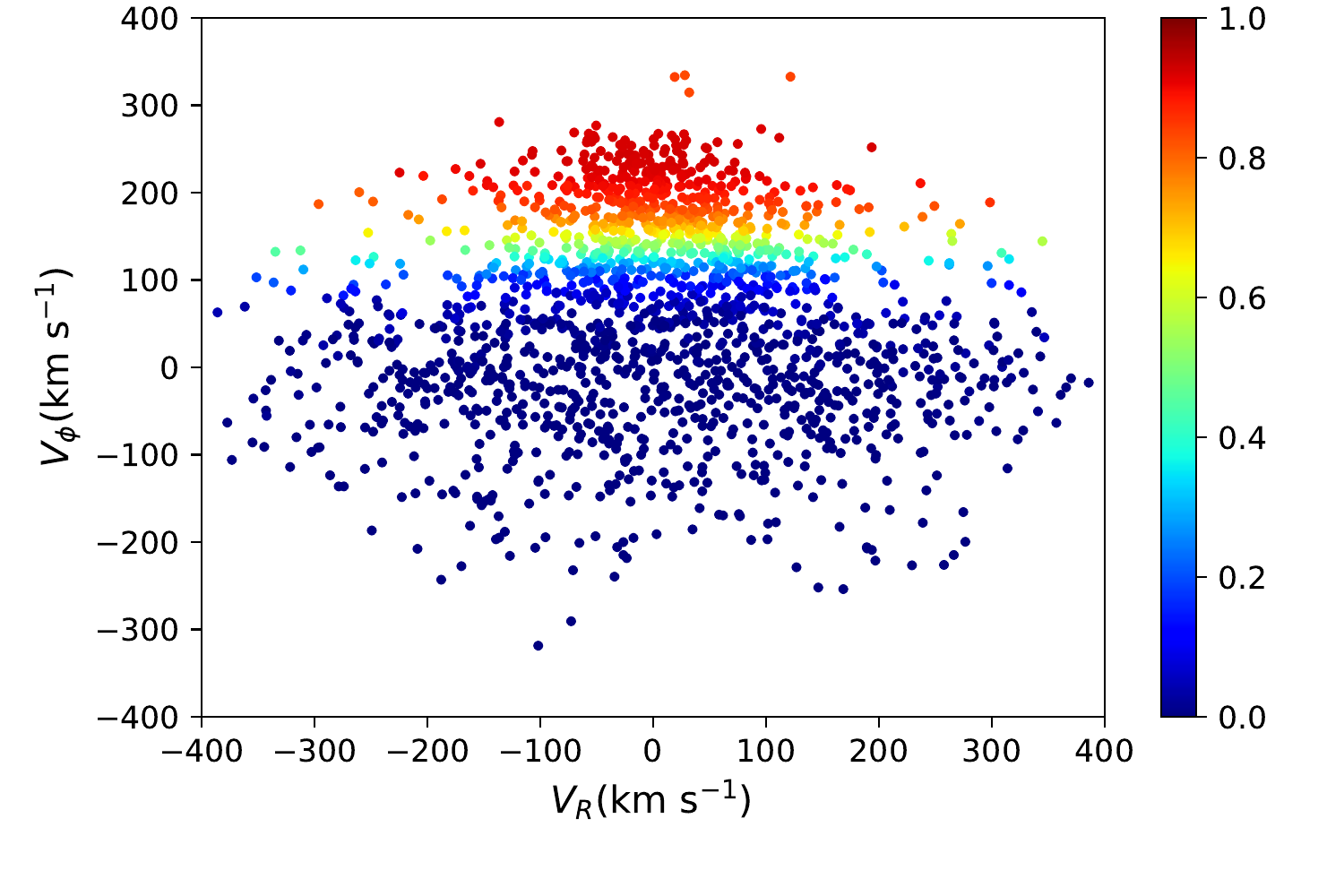}\\ \vspace{0.0cm}
    \caption{ The phase space distributions of the stars in the different volumes from S-sample are shown.
    The left and right panels show the distribution in action $J_{\phi}$ versus energy $E$ and velocity spaces,
    respectively. The stars are color-coded by the probability of belonging to the disk component for the volumes
    with $Z<4$ kpc, or the probability of belonging to the GES component  for the volume with $4<Z<6$ kpc.
    The dashed lines in the left panels represent the circular orbits. 
    From top to the bottom, the panels represent the volumes with different height, 
    e.g. $4<Z<6$ kpc,  $2<Z<4$ kpc, $1<Z<2$ kpc, $-1<Z<1$ kpc, respectively.}
    \label{fig:Phase_S}
\end{figure}

\begin{figure}
\hspace{0.01cm}
    \includegraphics[width=0.42\textwidth]{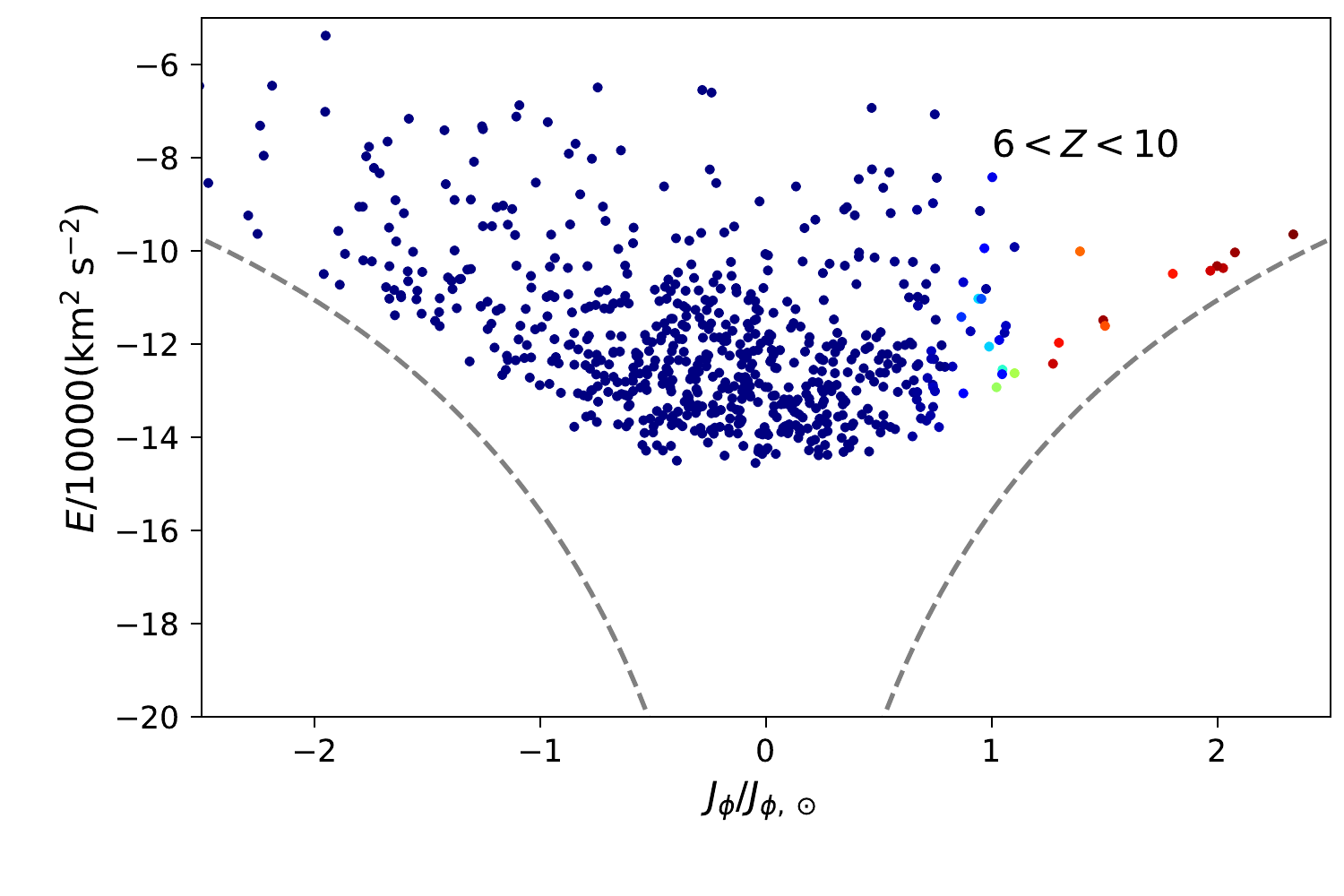} 
    \includegraphics[width=0.42\textwidth]{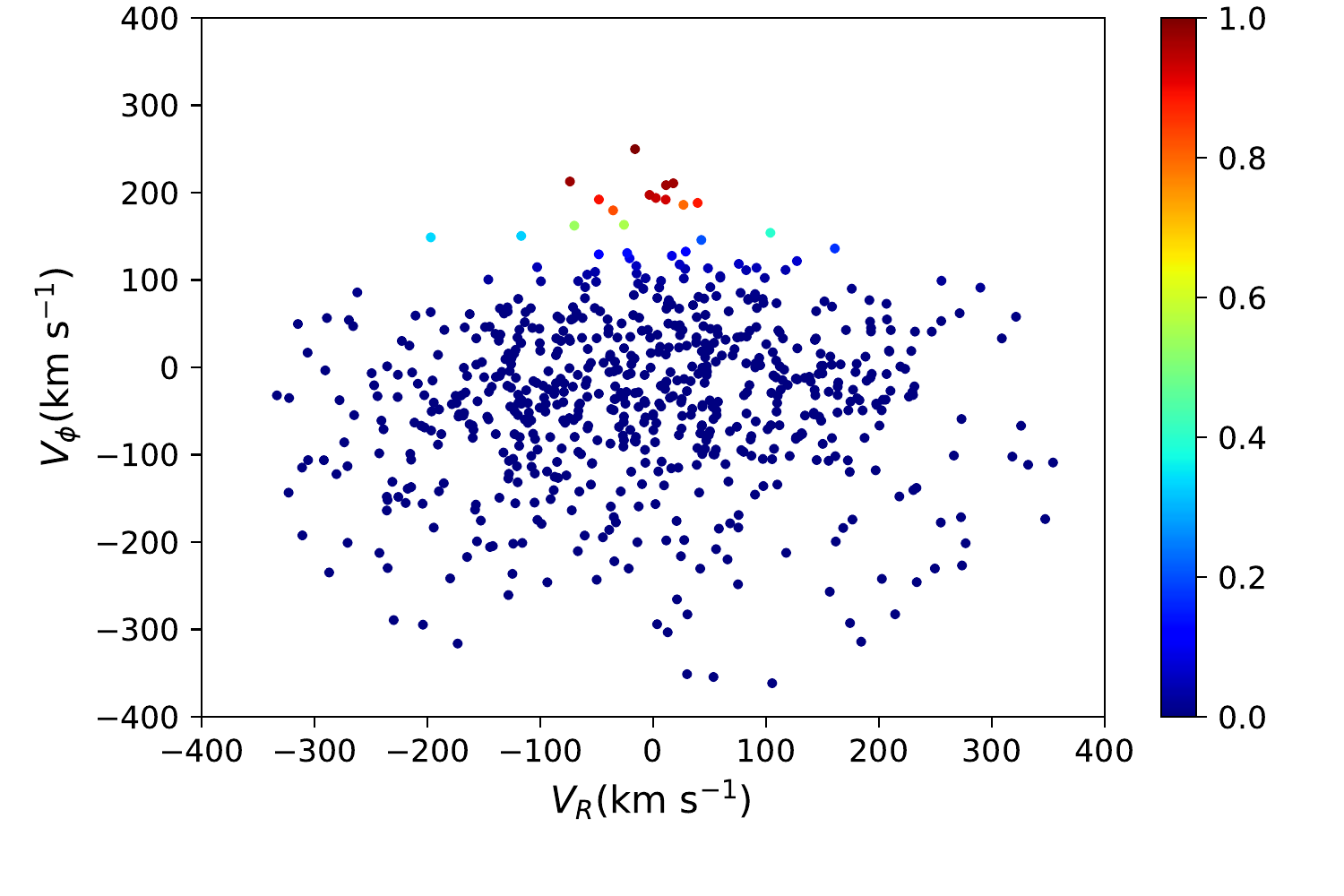} \vspace{-1.0cm}\\\vspace{-1.0cm}
    \includegraphics[width=0.42\textwidth]{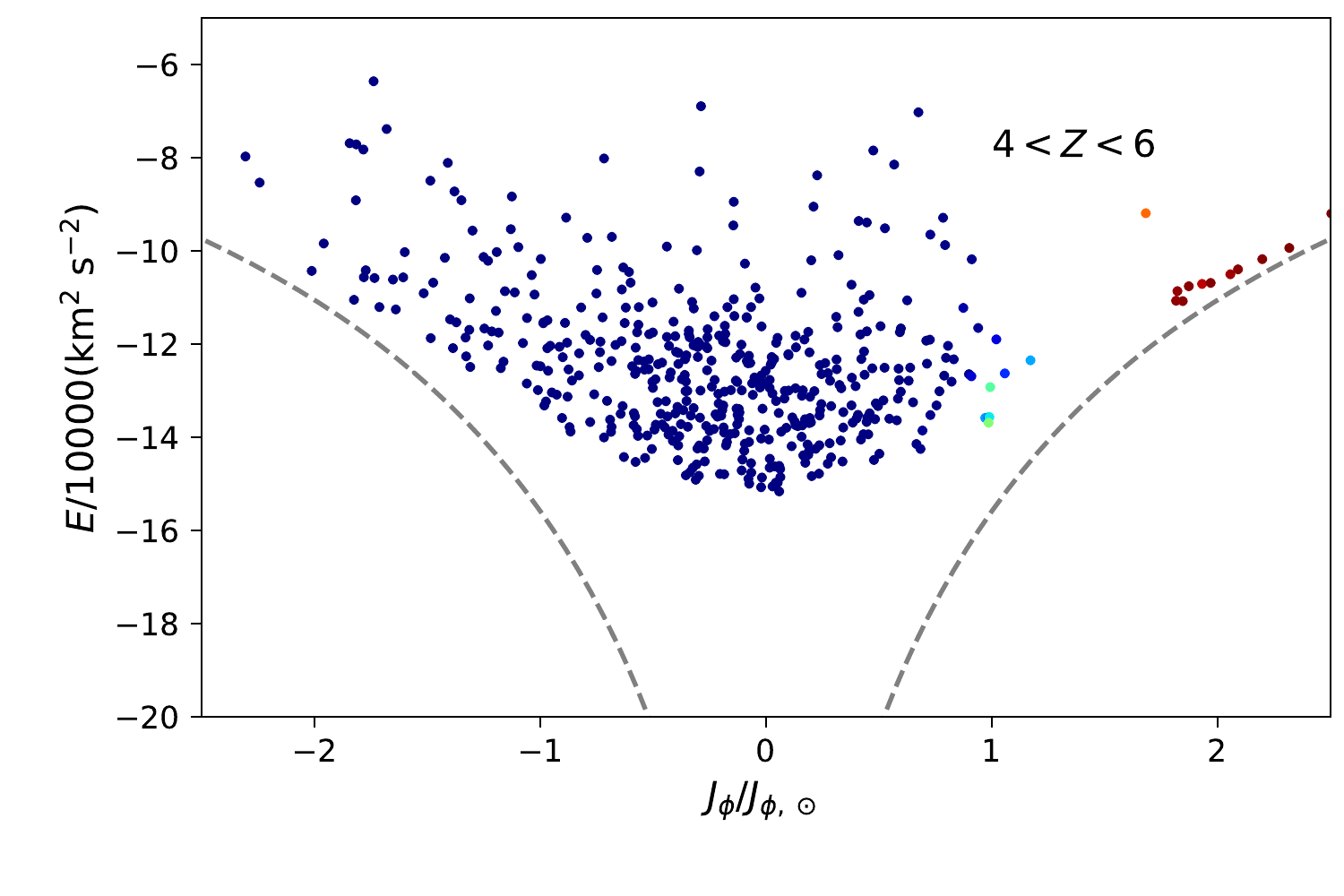}
    \includegraphics[width=0.42\textwidth]{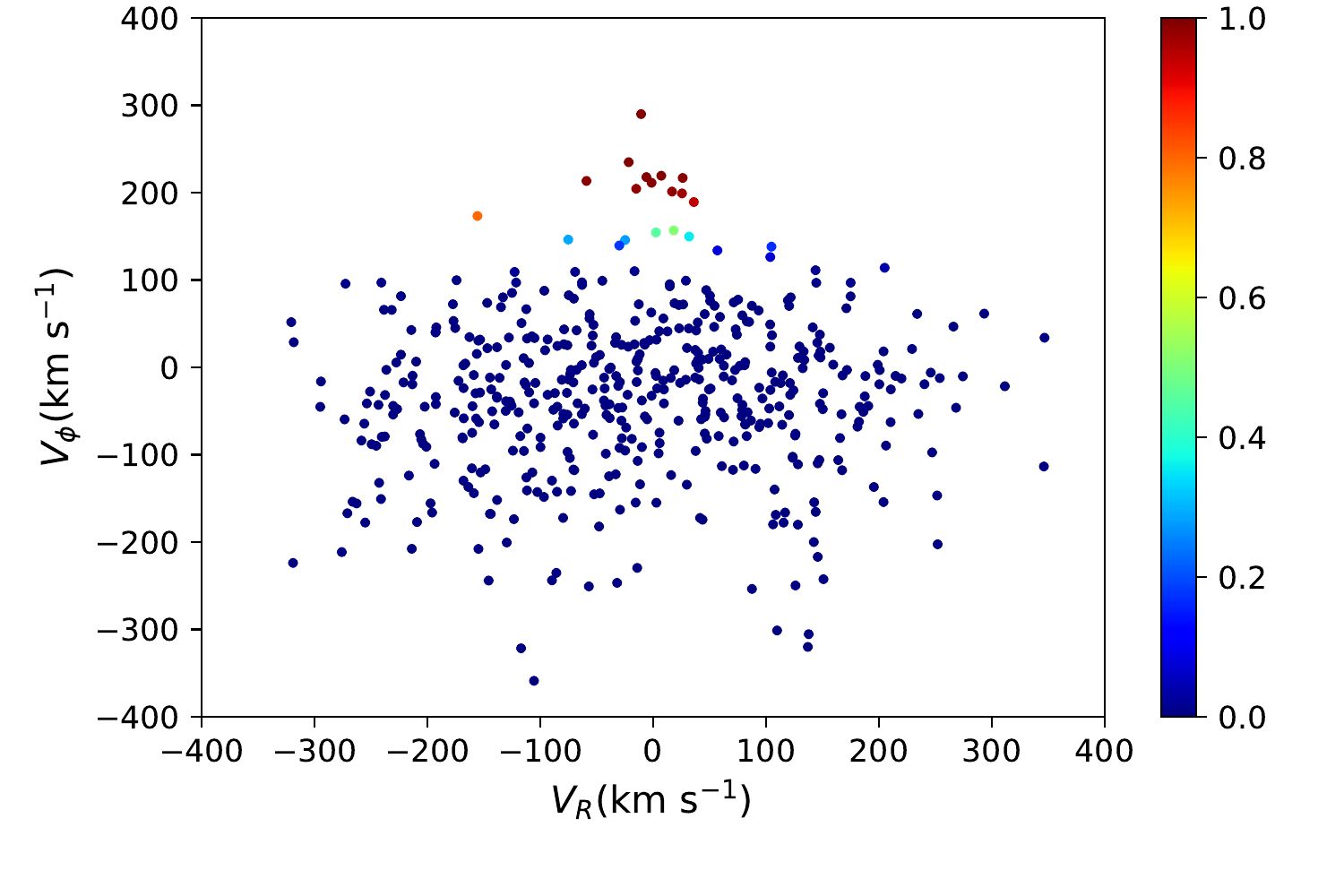} \\\vspace{-1.0cm}
    \includegraphics[width=0.42\textwidth]{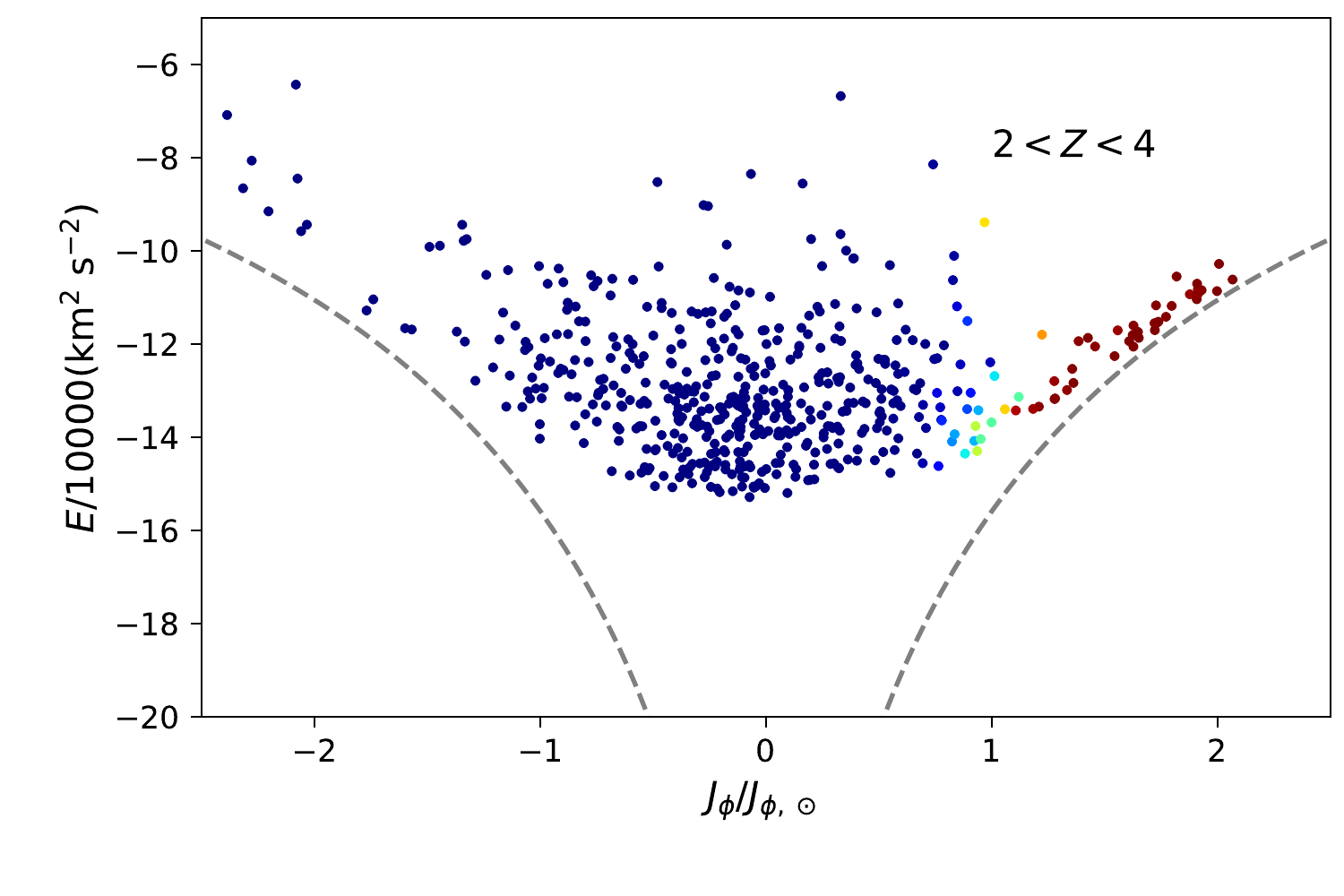}
    \includegraphics[width=0.42\textwidth]{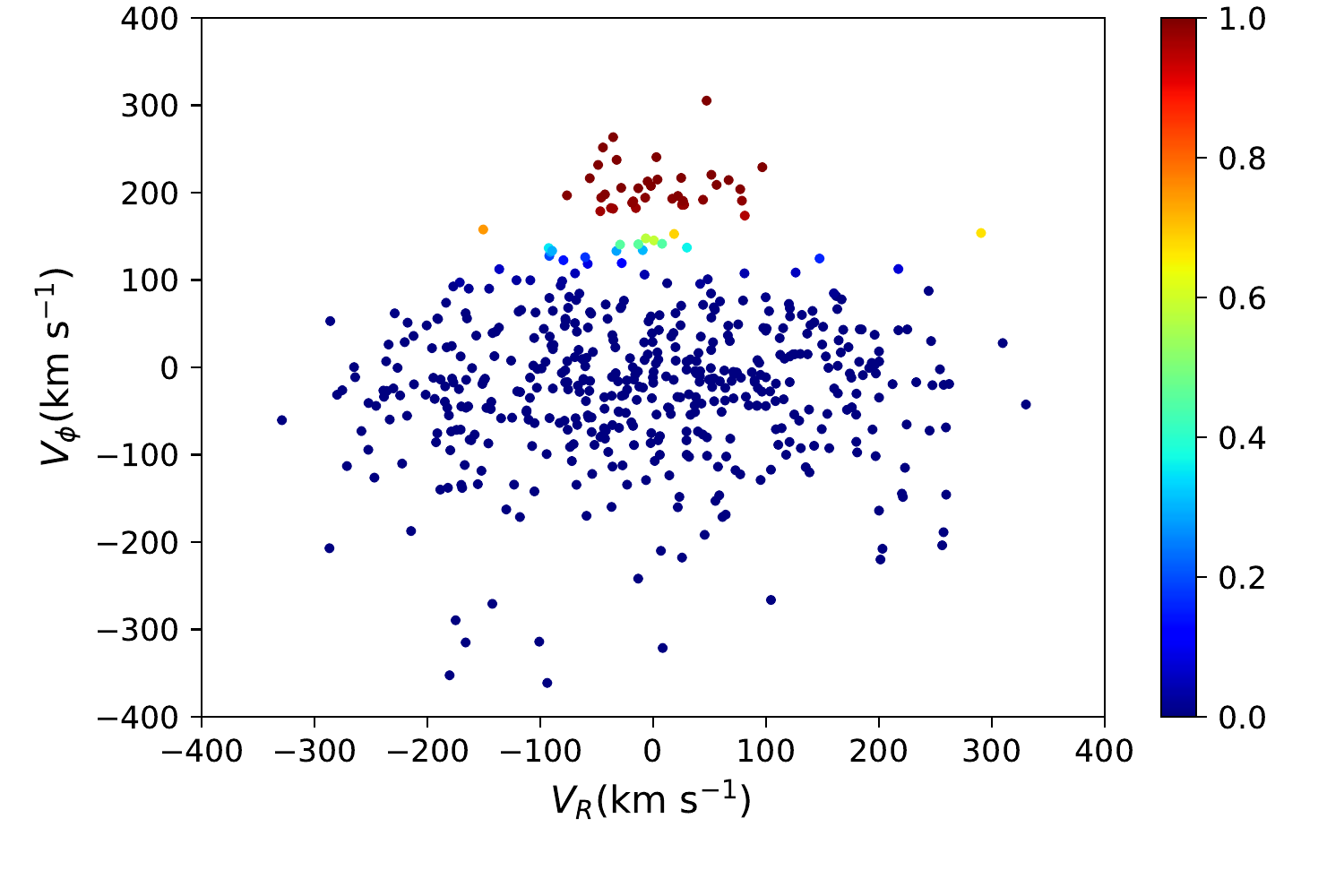}\\\vspace{-1.0cm}
    \includegraphics[width=0.42\textwidth]{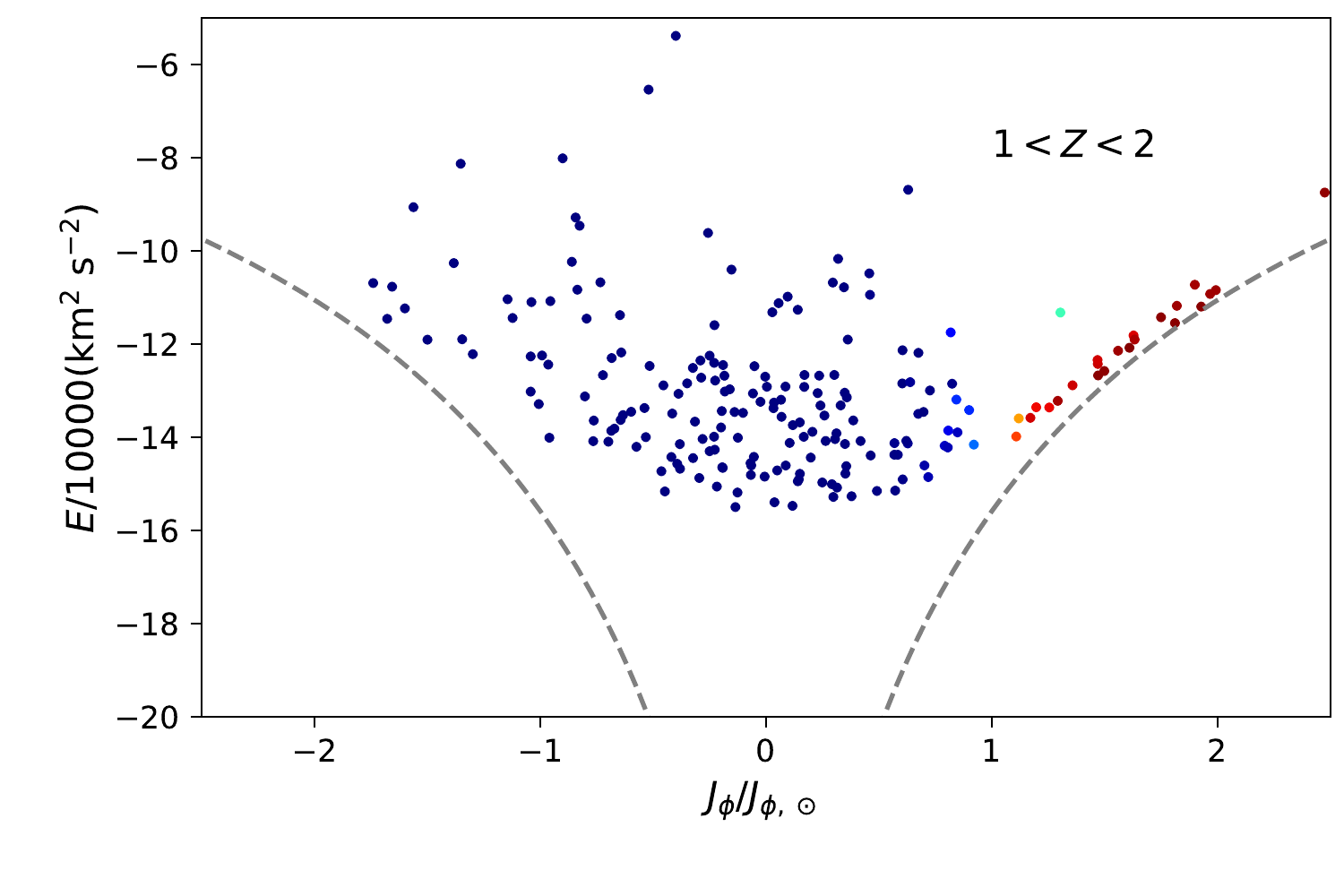}
    \includegraphics[width=0.42\textwidth]{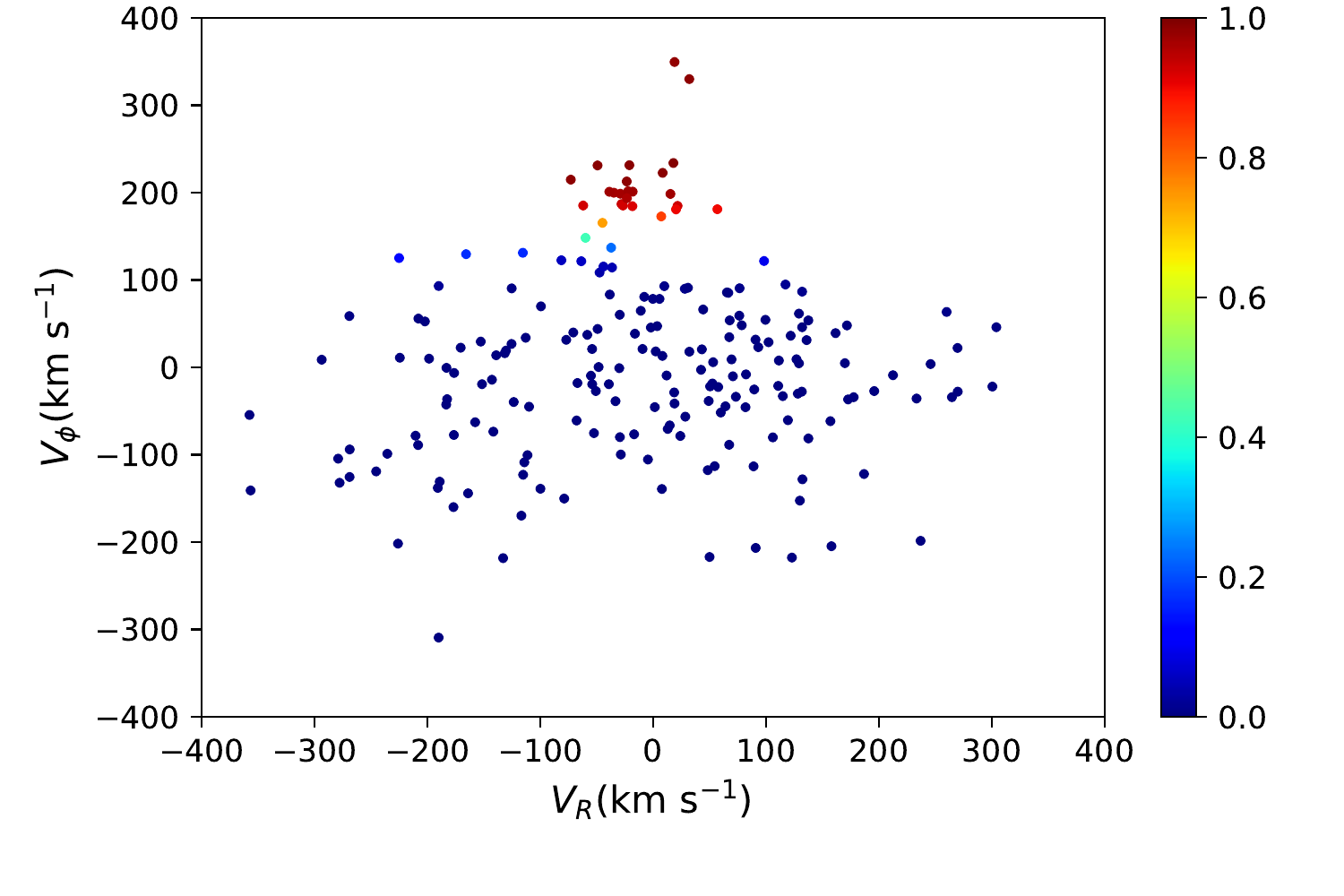} \\\vspace{-1.0cm}
    \includegraphics[width=0.42\textwidth]{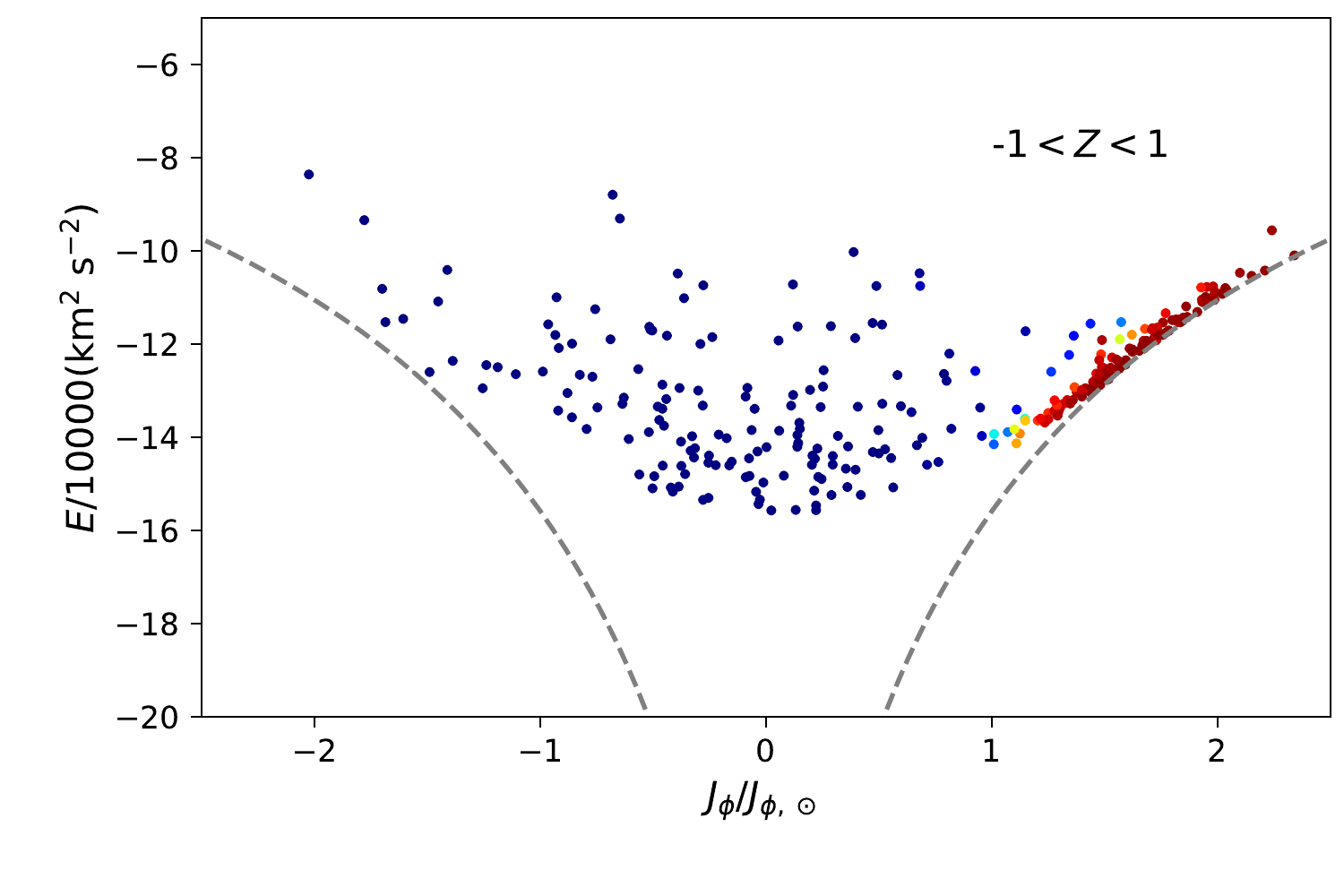}
    \includegraphics[width=0.42\textwidth]{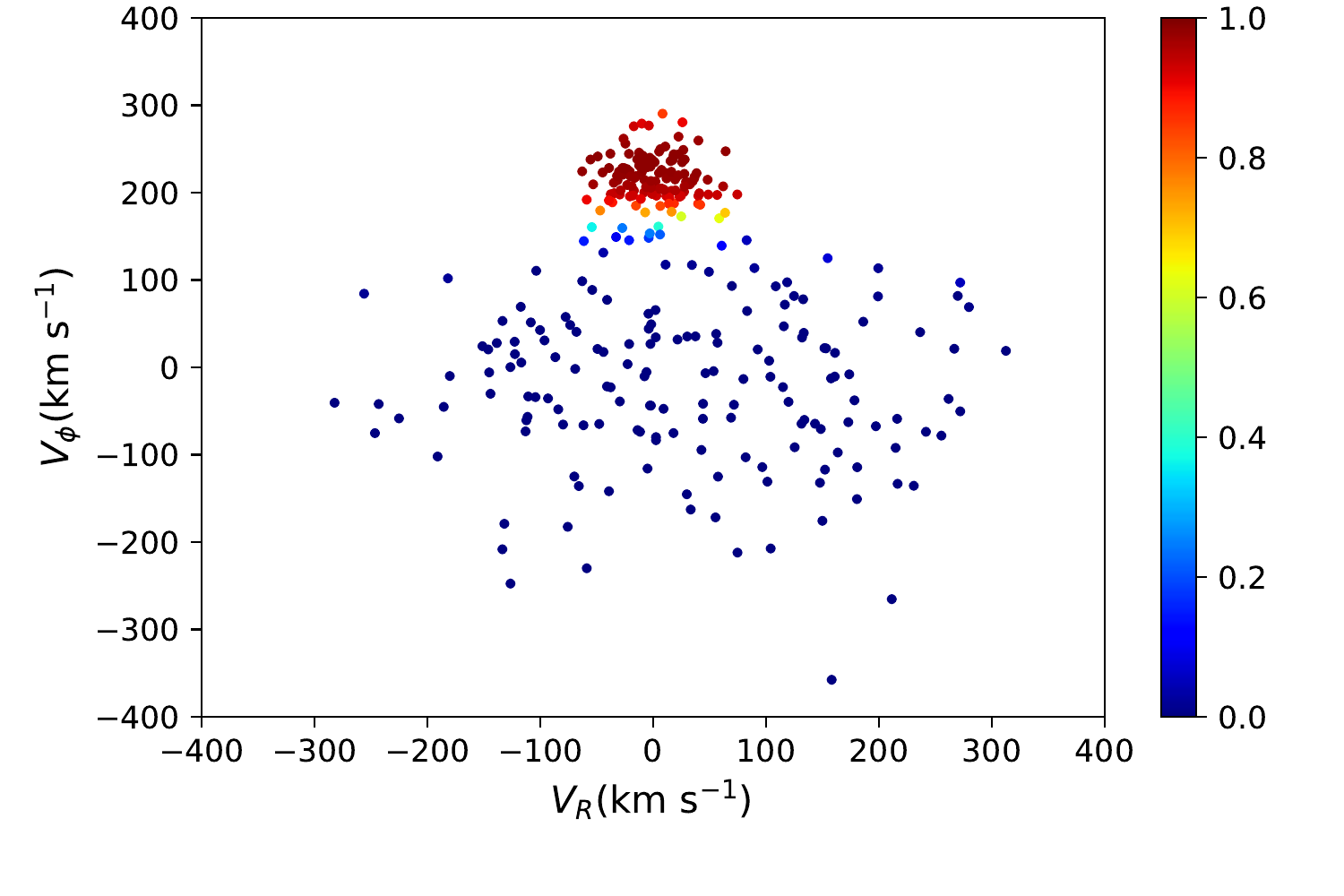} \\ \vspace{0.0cm}
    \caption{ Similar to Figure~\ref{fig:Phase_S}, but for the stars from SO-sample.
    From top to the bottom, the panels represent the volumes with different heights, 
    i.g. $6<Z<10$ kpc, $4<Z<6$ kpc, $2<Z<4$ kpc, $1<Z<2$ kpc, $-1<Z<1$ kpc, respectively.
    The dots are color-coded by the probability of belonging to the disk component.}
    \label{fig:Phase_SO}
\end{figure}

 According to the distribution of the components, we calculate the 
probabilities of the components for each star. 
In this way, Figures~\ref{fig:Phase_S} and \ref{fig:Phase_SO} 
show the distributions of the stars in each volume color-coded by the probability to the disk component.
Because the disk component is almost vanished in the higher volumes in S-sample,
where there are only the halo and the GES components,
we represent the probability of the GES in the subsample with $4<Z<6$ kpc (the top left 
panel in Figure~\ref{fig:Phase_S}). 

\begin{figure}
 \centering
\hspace{0.01cm}
    \includegraphics[width=0.82\textwidth]{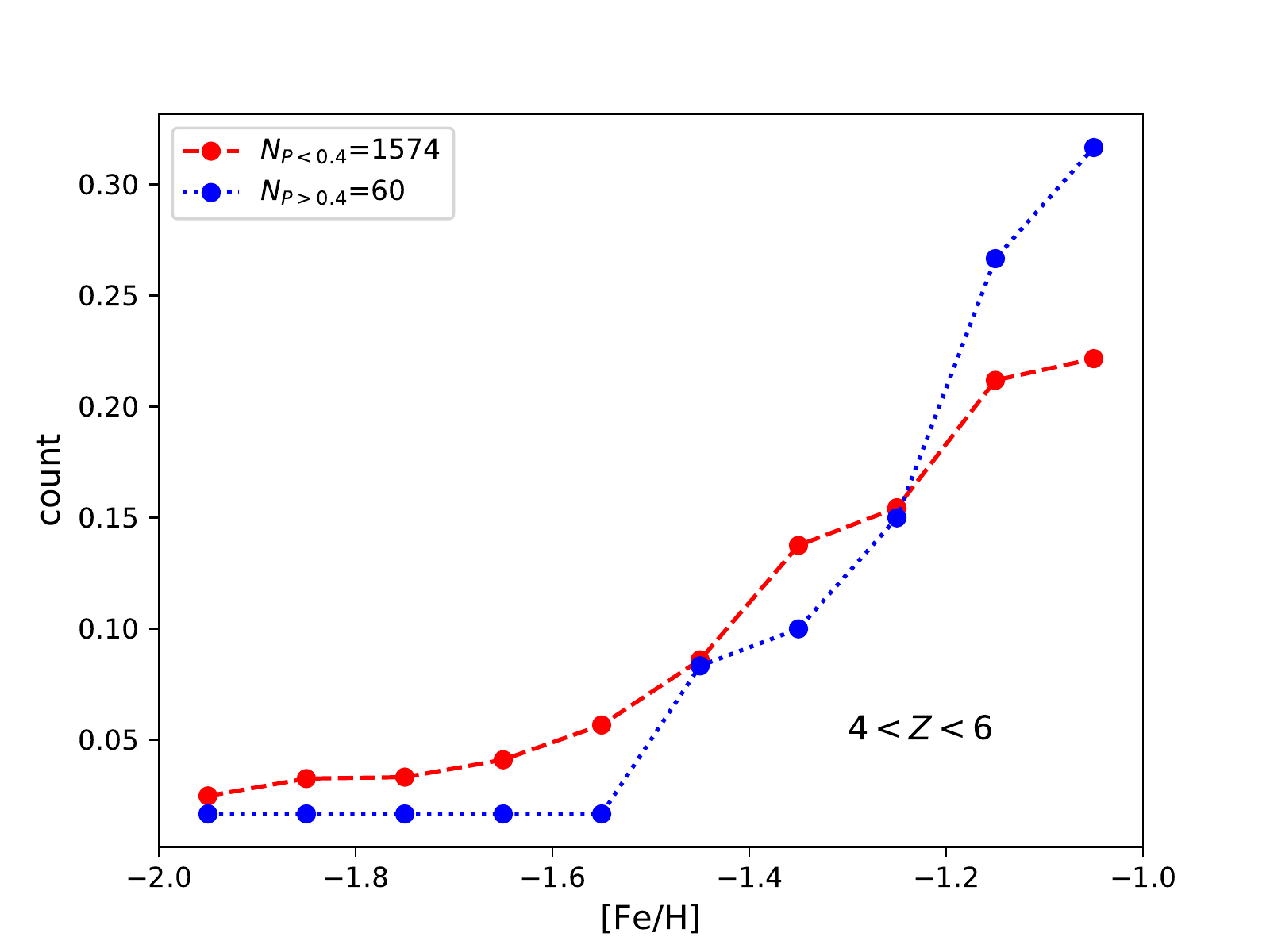} 
    \caption{ The histogram distribution of the metallicity of the stars with in the S-sample with
     $4<Z<6$ kpc.  The distributions of the stars with probability higher and lower than 0.5 are represented 
     with blue and red symbols, respectively.}
    \label{fig:GES_FEH}
\end{figure} 

 GES is included in the Bayesian model as an independent component, 
because it has significantly different 
dynamical information with the halo \citep{2018Natur.563...85H}, smaller  rotational
velocity and dispersion and relatively larger energy. As showed in
Figures~\ref{fig:Phase_S} and \ref{fig:Phase_SO}, the probability for the GES member stars are 
around 50\% at most. That means it is quite difficult to select a pure sample to study its 
chemical information. In Figure~\ref{fig:GES_FEH}, the metallicity distributions of the stars 
with probability higher or lower than 0.4 in S-sample with $4<Z<6$ kpc are showed. We 
find that the distribution with higher probabilities is still 
different with that from \cite{2018Natur.563...85H}, because of the high contamination ($>50\%$).

What should be noticed is that the GES component is more significant in the 
top right panel in Figure~\ref{fig:Phase_S}, where the stars are located with
$4<Z<6$ kpc, but this is not clear for 
the outer volumes with $12<R<20$ kpc in Figure~\ref{fig:Phase_SO}.  It is not clear that
the \emph{missing} of GES is intrinsic or that the larger uncertainties of the distances make 
the distribution of the action $J_\phi$ more diffused.

\subsection{Rotational velocity distribution of the disk}
 Besides the halo component, the variances of
the rotational velocity and its dispersion of the disk component are also 
obtained with the sample [Fe/H]$<-1$, even though this may not represent the whole typical thick disk. 
What should be noticed is that our model for the Bayesian method does not include the 
 metal weak thick disk \citep{Carollo2019ApJ...887...22C} 
as a independent component, because of 
the low fraction in our sample.

From the top panels in Figure~\ref{Fig:variance} which represent
the results of S-sample, we  find that the rotational velocity 
of the disk component (red symbols) decreases with increasing 
height to the disk plane, while the dispersion increases.
As those stars with 
metallicity [Fe/H]$>-1$ are removed, there are too few thin disk stars 
are left in our sample \citep{2015ApJ...808..132H} to bring a significant offset.
From Table~\ref{tab:MCMC_results}, we  find that the rotational
velocity is around 185$^{+5}_{-5}$ km s$^{-1}$ for the volume $-1<Z<1$ kpc, which
also suggests that it is contributed by the thick disk component \citep{1990AJ....100.1191M}.
It decreases to 127 km s$^{-1}$ of the volume with $2<Z<4$ kpc for the S-sample. Meanwhile,
the dispersion of the rotational velocity increases from  41 km s$^{-1}$ at lowest volume to 55 km s$^{-1}$.
The decreasing trend of the rotational velocity and the increasing trend of the rotational 
velocity dispersion support the conclusion of \cite{2012MNRAS.425.2144L}, 
that there may be two components of the thick disk.

Different with the results from the S-sample, the rotational velocity of
the disk component in the SO-sample decreases firstly and becomes  flat, around 200 km s$^{-1}$, at higher $Z$.
The rotational velocity dispersion for SO-sample increases 
with the height to the disk plane  from 20 km s$^{-1}$ to 45 km s$^{-1}$. 

Figures~\ref{fig:Phase_S} and \ref{fig:Phase_SO} show the distribution of
the  stars in $E-J_{\phi}$ and $V_\phi-V_R$ spaces, respectively. 
The stars are color-coded by the probability of belonging
to the disk or the GES. We 
find that the disk stars with high probabilities (the red dots) in the SO-sample are much closer to
the circular orbit line (the dashed line) than those in the S-sample.
    Comparing with the results from \cite{2012ApJ...757..151L} and \cite{2015ApJ...801..105X},
those  stars of high probabilities to be disk members with rotational velocity of $\sim200$ km s$^{-1}$ overlap
with the substructure Monoceros Ring. 
As claimed by \cite{2015ApJ...801..105X}, the disk stars can be heated by the disk oscillations and
reach to larger heights. So in this work, those stars with larger rotational 
velocities are possible the extension
of the disk, which is also approved by Li et al (2020, in prep.), who analyse those
Galactic Anticenter Substructures including the Monoceros Ring and 
the Triangulum-Andromeda cloud in dynamical  and chemical spaces.

\subsection{Disk Flare}

The red symbols in the bottom panels of Figure~\ref{Fig:variance} show the 
distributions of the disk for outer volumes with $12<R<20$ kpc.
The green symbols represent the results with an additional component in the model. 
According to the rotational velocity and its dispersion, this component is an extension of the disk.
In other words, the disk component extends to higher volumes up to
6$\sim$10 kpc with galactocenteric distance
$R$ between 12 and 20 kpc. This is also represented by the red symbols in Figure~\ref{fig:Phase_SO}. 
That is the disk flare, that the outer disk is much thicker.  This is also indicated
with scale length distribution by \cite{2018MNRAS.478.3367W} using the same K giant sample. 

\section{Discussion}\label{Sec:Disc}

\subsection{Interaction between the halo and the disk}\label{Sec:Disc:interaction}
\begin{figure*}
\gridline{\fig{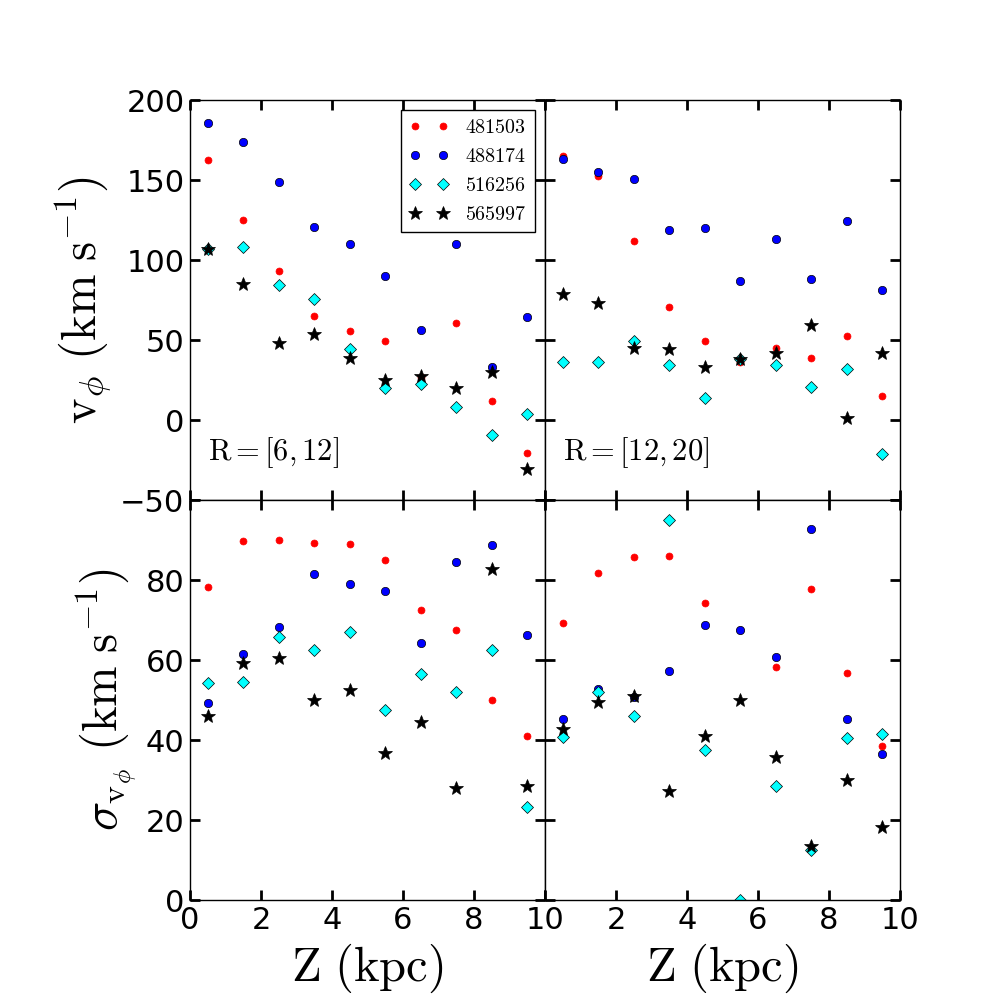}{0.48\textwidth}{(a) star a}
                  \fig{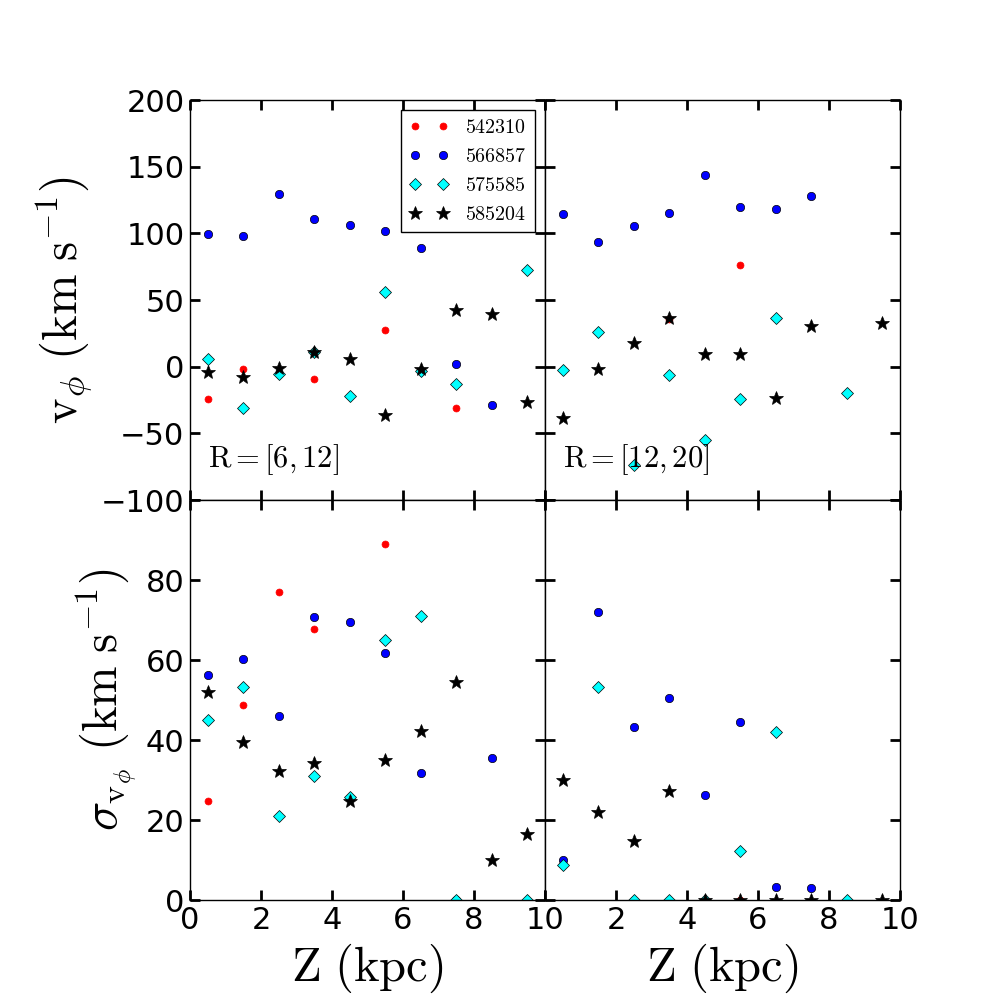}{0.48\textwidth}{(b) star b}
          }  \vspace{-0.3cm}
\caption{The velocity $v_{\phi}$ and velocity dispersion $\sigma_{v_{\phi}}$ 
distributions for eight galaxies.  (a) the results of star particles for four 
galaxies, the axis ratios from the star particles for galaxy 481503, 488174, 
516256 and 565997 are 1:0.997:0.213, 1:0.988:0.312, 1:0.988:0.432 and 1:0.995:0.325, 
respectively. 
(b) Same as panel (a), the results for galaxies 542310, 566857, 575585 and 585204. 
The axis ratio from the star particles for these galaxies are: 1:0.985:0.751, 
1:0.982:0.749, 1:0.985:0.781 and 1:0.985:0.808, respectively. 
\label{fig:vphi_sim}}
\end{figure*}

There are many mechanisms to generate the differential rotation of the halo. 
One possible mechanism for the decreasing trend of the halo 
rotational velocity versus the height to the disk 
plane is the interaction between the halo and the disk. 
In order to check this 
scenario, we use the simulated galaxies from the TNG100 
simulation~\citep{2018MNRAS.475..676S,2018MNRAS.475..648P,2018MNRAS.480.5113M,2018MNRAS.477.1206N}. 
The TNG 100 simulation is a magnetohydrodynamic cosmological simulation, 
which contains $2\times1820^3$ resolution elements in a cosmological
in a ~$(\rm 110 Mpc)^3$ box.  
Compared with the original Illustris 
simulation~\citep{2013MNRAS.436.3031V,2014MNRAS.438.1985T}, 
the TNG simulation has adopted new physics models and improved 
the implementations of galactic winds, stellar evolution, 
chemical enrichment~\citep{2018MNRAS.473.4077P}, and AGN 
feedback~\citep{2017MNRAS.465.3291W}.  Therefore, the TNG 
simulation can reproduced many observed galaxy properties 
better and scaling relations to different degrees.  
Galaxies in their host dark-matter halos were identified using the 
\textsc{subfind} halo finding algorithm~\citep{2009MNRAS.399..497D}.  
In Figure~\ref{fig:vphi_sim}, we show the velocity $v_{\phi}$ and 
velocity dispersion $\sigma_{v_\phi}$ for eight galaxies from the 
TNG simulation with different axis ratios.  It is  seen that there are clear velocity gradients 
in $v_{\phi} $ for stars  in the panel (a). For these four galaxies, 
they exist the oblate-disk, with the axis ratio $c/a$ (minor axis
over major axis) lower than 0.5. 
For the other four galaxies as shown in panels (b) , the 
stellar systems are nearly spherical or triaxial, with $c/a>0.5$,
there are no obvious velocity gradients in $v_{\phi}$ versus height to the disk.

As shown in~\cite{2019MNRAS.483.3048W}, the oblate galaxies have 
larger spin parameters than prolate and triaxial galaxies. 
In other words, the oblate system has the larger rotational velocity. 
Close to the oblate-disk,  the fast rotation disk dominates the 
rotation velocity. With the height increasing, the 
halo begins to dominate the kinematics of the system, and the 
halo is more spherical or triaxial.  Therefore,  
 $v_{\phi}$ decreases with the height to the disk.   In other words, the decreasing
trend of the halo versus the height to the disk is quite likely caused 
by the dynamical interaction between the 
disk and the halo. A stronger disk makes the larger decreasing rate. From the comparison,
we confirm that the halo must be oblate, which is consistent
with the conclusion of \cite{2018MNRAS.473.1244X}. 

\subsection{Interaction between the halo and the bar}\label{Sec:Disc:Disk_bar}
Our Milky Way is a typical barred galaxy and 
the bar can affect the redistribution of the angular momentum in the system.
Angular momentum is emitted from the bar region and absorbed by the 
corotation resonance (hereafter CR) and outer Lindblad resonance (hereafter OLR)
in the disk, and also absorbed in the spheroid components by all resonances \citep{2013seg..book..305A}.
The pattern speed of the Milky Way bar is 
$40\sim60$ $\rm {km\ s^{-1}\ kpc^{-1}}$ \citep{2012MNRAS.427.1429W, 2013MNRAS.435.3437W, 
2013MNRAS.428.3478L,Portail2017MNRAS.465.1621P} , 
and the corresponding OLR radius is smaller than 8.5 kpc.  
In our SO-sample, we still find the clear rotational trend for the halo star, 
therefore, the effect from the bar is small for our findings here.
In the other hand, the disk component can extend to the outer part, even as far as 20 kpc. This suggests the 
possibility for the disk to affect the halo spin \citep{2012MNRAS.419.1951V}.
\subsection{The halo assemble history }\label{Sec:Disc:haloAss}
The third possibility for the rotation of the halo is the
assemble history. Those merged satellites should have a random angular momentum
distribution \citep{2015ApJ...801...98S}, unless that most of those satellites fall in  groups and those
groups dominating the inner halo, such as the 
GES \citep{2018Natur.563...85H, Belokurov2018MNRAS.478..611B} and \emph{Sequoia} \citep{2019MNRAS.488.1235M}.
As we discussed above, we treat the GES as a different component in the model. Meanwhile, the
\emph{Sequoia} has a very retrograde rotational velocity. 
The possibility is too low to lead a decreasing trend for the rotational velocity versus the height. 

\subsection{The dichotomy of the halo} 
\label{Sec:Disc:Dich}
 Another possible 
explanation for the decreasing trend is the dichotomy of the halo. 
 \cite{Carollo2007Natur.450.1020C} found that the inner  and the outer part 
of the halo have different chemical and dynamic properties. The dichotomy were 
confirmed by \cite{Fernadez2015A&A...577A..81F} and
\cite{Yoon2018ApJ...861..146Y} with abundance distributions of calcium, magnesium and carbon.
According to the results of conclusion of \cite{Carollo2007Natur.450.1020C}, 
the inner halo rotates progradely
with a modest speed. In contrast, the outer part is retrogradely rotating. 
 \cite{An2013ApJ...763...65A} also found the similar results, 
that the retrogradely rotating stars are generally more metal poor.

Considering the large overlapping of the inner 
and outer halo and their different rotation behavior \cite{Carollo2007Natur.450.1020C}, 
it is  possible to 
bring a decreasing trend 
of the rotational velocity versus the height to 
the disk plane in the transition region of the two halo. 
As the contribution of the inner halo 
decreases with a higher volume and the rotational velocity of the complex
will decrease. Surprisingly, as showed in Figure~\ref{fig:VT_Z_fit}, 
the trend in the S-sample is significantly steeper
than that in the SO-sample. This suggests that the decreasing trend in the SO-sample may be caused
by the dichotomy of the halo, but that in the S-sample is not 
mainly caused by the dichotomy of the halo, at least this may not be the main reason. This is also supported
by  the rotational velocity 
dispersion distribution. As showed in Figure~\ref{Fig:variance}, the dispersion  in 
the S-sample is almost flat, which suggests that
the inner halo is dominating. Meanwhile the variance of the rotational velocity 
dispersion is changing significantly for
the SO-sample. It indicates that the dichotomy of the halo plays an important role to generate
the decreasing trend in the SO-sample.

Above all, we claim that the dichotomy of the halo plays an important
role in generating the decreasing trend of the rotational velocity
distribution in the SO-sample, but this is not the main mechanism for the decreasing trend in the S-sample.

\subsection{Effect of the distance calculation}\label{Sec:Disc:Dist}

To figure out if the rotational velocity distribution will be affected by the distance calculation,
we firstly  redo the procedures with different distance correction
coefficients. Based on the previous distance correction, we multiply an 
additional coefficient to the corrected distance, e.g. 0.9 and 1.1. After all the same steps,
we find that, the rotational velocity decreasing trend for the halo component 
is still there, but the decreasing rate (the slope) varies slightly, 
from $-3.03\pm0.73$ km s$^{-1}$ kpc$^{-1}$ to $-3.79\pm0.97$ km s$^{-1}$ kpc$^{-1}$ for inner volumes with coefficients
of 0.9 and 1.1, respectively. Meanwhile for the outer volumes the decreasing rate varies from
$-2.06\pm0.46$ km s$^{-1}$ kpc$^{-1}$ to $-1.30\pm0.37$ km s$^{-1}$ kpc$^{-1}$ with the coefficients of 0.9 and 1.1, respectively.
That means the rotational velocity gradient is intrinsic, and the 
gradient for the outer volume is  a bit shallower.

\section{Summary}\label{Sec:Sum}
We use the K-giant stars from LAMOST DR5 to investigate the rotation information of the halo and the disk. 
We find that the rotational velocity of the halo decreases with an increasing
height to the disk plane. The dispersion of the halo 
is almost flat up to 15 kpc. The rotational velocity of the inner part decreases 
 faster than that of the outer part, $-2.75$ and $-1.88$  km s$^{-1}$ kpc$^{-1}$respectively. 
 Analysing all the possible mechanisms for the decreasing trend,
 we claim that the decreasing trend suggests an oblate halo profile, 
which is consistent with that revealed by \cite{2018MNRAS.473.1244X}. 
This is possibly caused by the interaction between the halo and the disk component.

The signal of merging event GES is clear shown only in the volumes with height from
2 to 10 kpc and galactocentric distance between 6 and 12 kpc. Meanwhile  the rotational velocity dispersion
of the disk  is larger in  higher volumes. 
At the same time, the rotational velocity decreases versus the height.
Our results also show a flaring disk, which can reach the height of 6 to 10 kpc 
with galactocentric distance from 12 to 20 kpc. Limited by the sample, we 
claim that the disk can reach at least 20 kpc. 

 In order to avoid the contamination of different components, we use the Bayesian method to determine 
the rotational velocity for each component at different locations statistically. With the rotational velocity distribution
of each component,
it is able to determine the probability for each star belonging to the different components.  
As showed in Figures~\ref{fig:Phase_S} and \ref{fig:Phase_SO}, only the disk component
can be well picked out. 
The GES and the halo are still difficult to separate.
This will be improved 
in future for member stars selection of the two components using the full phase space information. 
More purer samples will greatly help the chemical studies in future with 
spectral data sets from LAMOST \citep{Liu2020arXiv200507210L},
SEGUE \citep{Yanny2009AJ....137.4377Y} and APOGEE \citep{Majewski2017AJ....154...94M}.

\appendix
\section{Distance correction}\label{sec:D_correct}
To make sure the distances from Gaia DR2 and LAMOST are consistent, 
we firstly select the common K-giant stars in both Gaia DR2 and LAMOST DR5 with following criteria,
\begin{itemize}
\item $\omega/\omega_{err}>5$
\item $D_{Gaia}<3$ kpc
\item $\sigma_{D_{Gaia}}/D_{Gaia}<0.1$
\item $SNR>10$
\end{itemize}
The first criterion is used to select those stars with parallaxes are well measured. The second and the 
third ones are used to constrain the distance $D_{Gaia}$ smaller than 3 kpc and high accuracy,
which is provided by \citet{2018AJ....156...58B}. 
The last item is used to constrain the data from LAMOST DR5
with high signal-to-noise ratio. After the selection, we have 5560 K-giant stars with distance accurately
measured by Gaia and reliable spectra from LAMOST DR5.

To compare the distances we define the difference $\Delta=\frac{D_L-D_{Gaia}}{D_{Gaia}}$, where $D_L$ is
the distance provided by \cite{2014ApJ...790..110L}, 
and $D_{Gaia}$ is the distance from \citet{2018AJ....156...58B}.
\begin{figure*}
    \centering
    \includegraphics[width=0.43\textwidth]{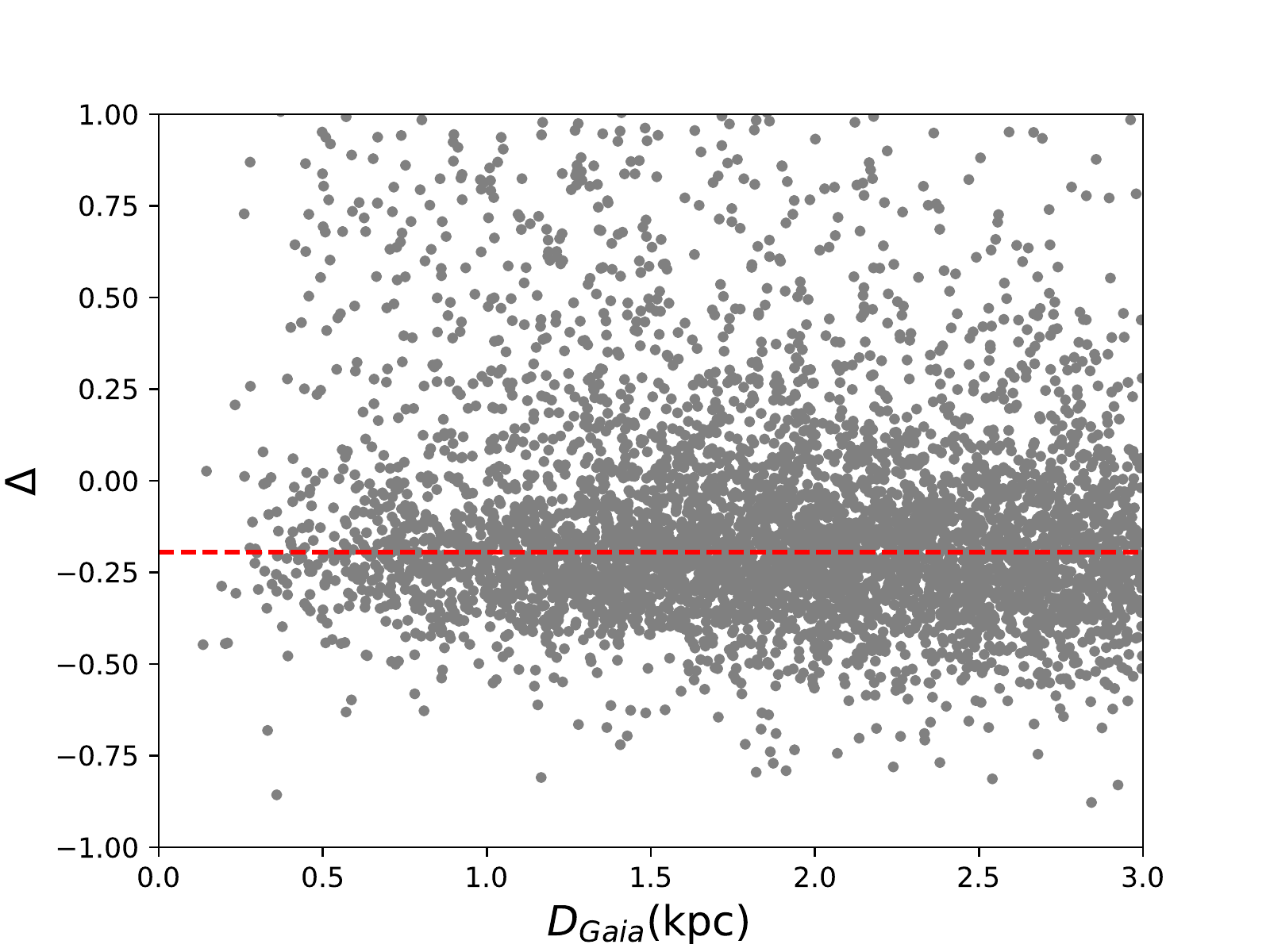}
    \includegraphics[width=0.43\textwidth]{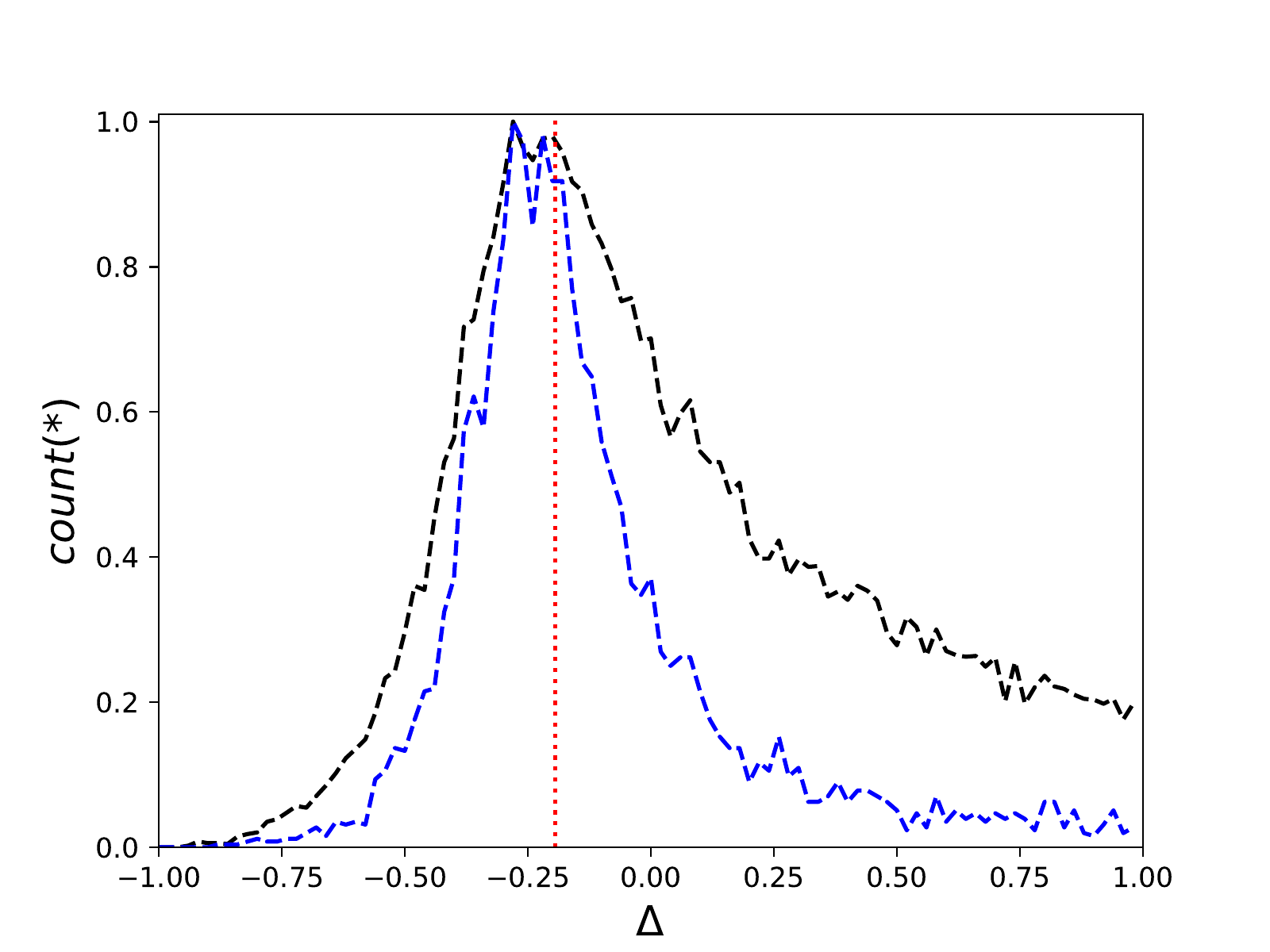}
    \caption{Left panel: The distribution of the distance difference $\Delta$ is shown versus the Gaia distance $D_{Gaia}$ with 
    the gray dots. Right panel: The histogram distribution of the distance difference $\Delta$  is shown. The black line represents
    the distribution of the whole sample while the blue line represents the distribution of the stars with reliable
    Gaia distances and high signal-to-noise ratios during LAMOST observation. The red lines in both panel represent
    the value for the distance correction.}
    \label{fig:Dist_comp}
\end{figure*}

\begin{figure}
    \centering
    \includegraphics[width=0.43\textwidth]{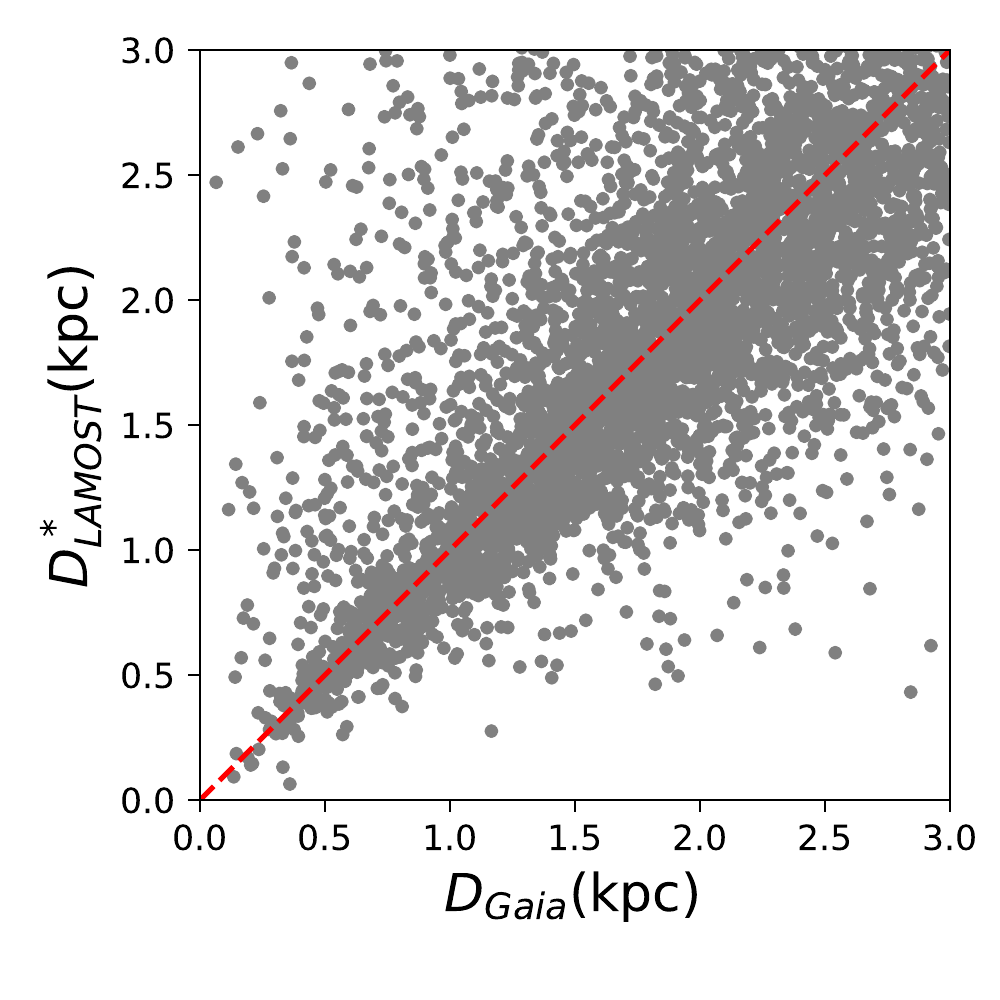}
    \caption{The relation between the corrected LAMOST distance $D_{LAMOST}^{*}$ and Gaia distance $D_{Gaia}$ is shown.
    The red line represents the distance ratio 1:1.}
    \label{fig:Dist_correct}
\end{figure}

Figure \ref{fig:Dist_comp} shows the distribution of the distance 
difference versus the Gaia distance $D_{Gaia}$
in the left panel and its histogram distribution in the right panel. 
From the distribution we can find that 
there is a system offset with $\Delta$ a constant around -0.2. To obtain the true value, we calculate
the mean value of $\Delta$ and its dispersion, $<\Delta>$ and $\sigma_{\Delta}$, 
and then select those stars within $1\sigma_{\Delta}$. 
Iterating this step until the mean value does not 
change significantly, lower than 0.005. The threshold is chosen 
because with this value the distance system 
error will be lower than 0.15 kpc at 30 kpc. Finally, we obtain the value  $\Delta\sim-$0.195, as shown
by the red line in both of the panels in Figure~\ref{fig:Dist_comp}. In this case,
the distance from LAMOST DR5 is corrected by dividing 0.805. 
Figure~\ref{fig:Dist_correct} shows the corrected LAMOST distance distribution as a function of the 
Gaia distance $D_{Gaia}$. The red line represents the 1:1 relation.

\section{Radial velocity correction}\label{sec:RV_correct}
The radial velocity provided by GDR2 is only available for bright stars with $G<14$. That will cause the sample 
cannot trace distant volumes. In this paper we adopt the radial velocity from LAMOST DR5. 
\begin{figure}
 \centering
 \includegraphics[width=0.46\textwidth]{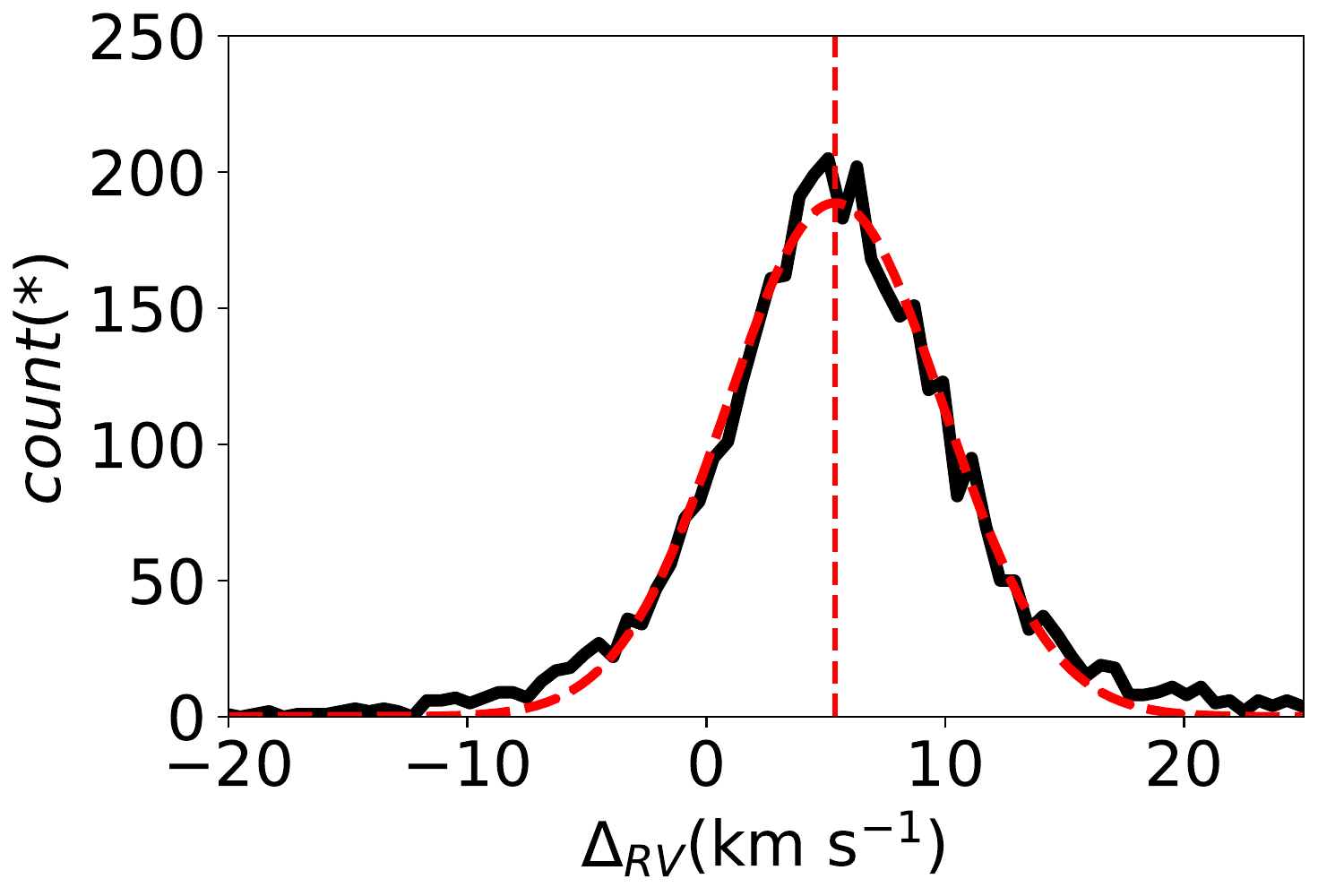}
 \caption{The radial velocity comparison between the values from LAMOST DR5 and GDR2. The red dashed line
 represent the fitting results with one Gaussian model, with mean value at $\sim$5.38 km s$^{-1}$ and 
 a dispersion 6.39 km s$^{-1}$.}
 \label{Fig:hist_rv}
\end{figure}
Figure \ref{Fig:hist_rv} shows the comparison between the radial velocities provided
by LAMOST DR5 and GDR2 of the common K-giant stars used in Paper I. We can clearly find an offset of
$\sim$5.38 km s$^{-1}$ and  a dispersion 6.39 km s$^{-1}$. In this paper, we correct the radial velocity
from LAMOST DR5 with +5.38km s$^{-1}$.

\section{Parameter determination in Bayesian method}\label{Sec:emcee}
\begin{figure}
\begin{center}
\hspace{0.01cm}
    \includegraphics[width=0.98\textwidth]{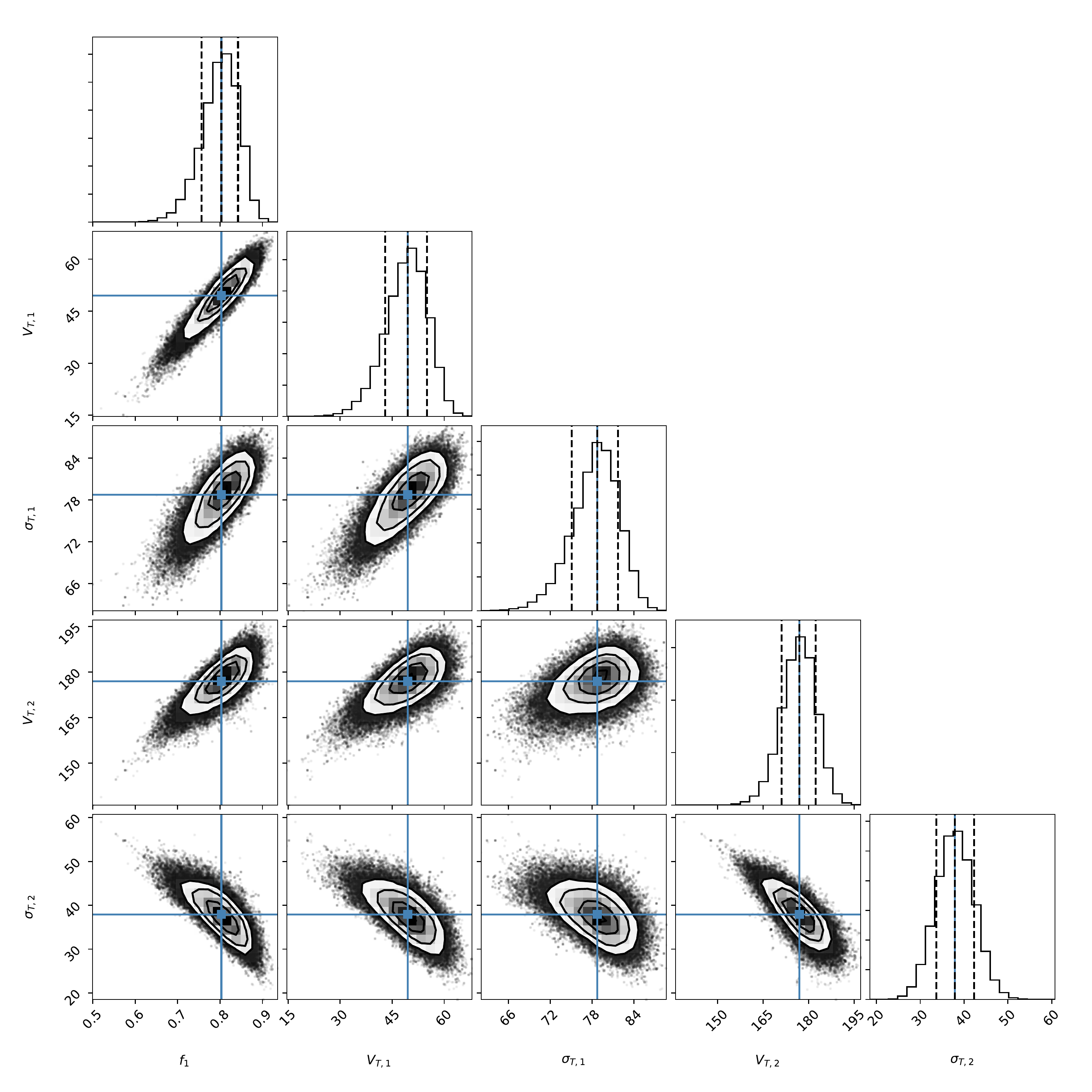}  
    \caption{Results from Bayesian method is showed for the volume with $1<Z<2$ kpc in S-sample.
    The dashed lines indicate the 16\%, 50\% and 84\% values for each parameter. The blue solid lines
    represent the median values for the parameters which are adopted for the components.}
    \label{fig:emcee_corner}
\end{center}
\end{figure}

\begin{figure}
\begin{center}
\hspace{0.01cm}
    \includegraphics[width=0.9\textwidth]{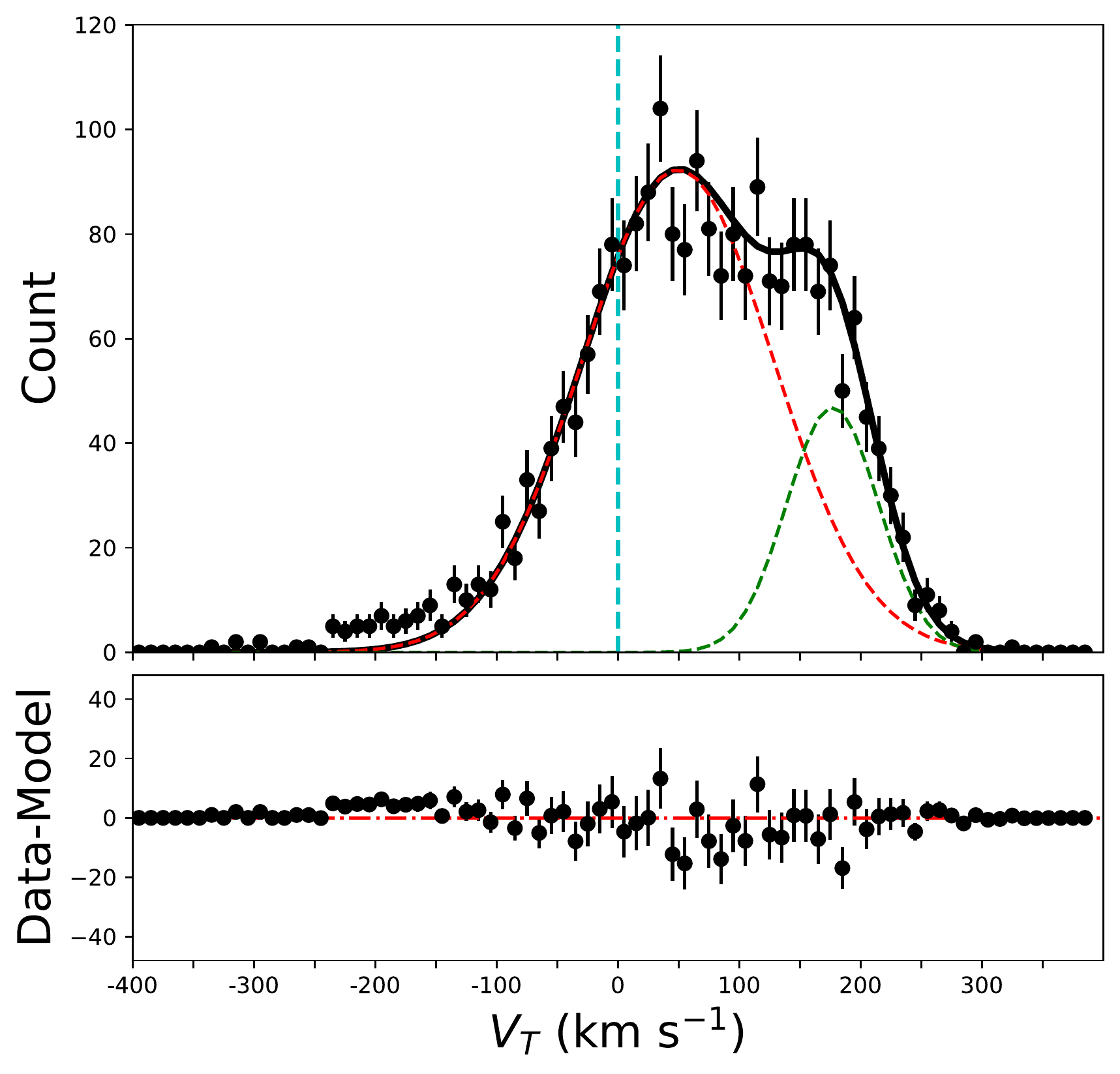}\\
    \caption{Top panel: The black dots indicate the histogram distribution of the rotational velocity. 
    The red and green dashed lines represent the distributions of the halo and disk components
    in the model defined during the  
    Bayesian method, respectively. The solid black line represents the distribution of all
    the stars constrained by the Bayesian method. The cyan vertical line indicates the value 0 for the rotational 
    velocity.
   Bottom panel: the residual distribution between the histogram distribution of the observational  rotational velocity and model.}
    \label{fig:emcee_fit}
\end{center}
\end{figure}

We adopt the median values for each parameter during the Bayesian method from \emph{emcee}. The 
upper and lower uncertainties are determined as the difference between the median value and the 84\% 
and 16\% values. Figure~\ref{fig:emcee_corner} shows the corner
distribution of the possible values for each parameter. The blue lines represent the median values 
for each parameter, while the dashed lines in each histogram distribution represent the 16\%, 50\% 
and 84\% values.
Figure~\ref{fig:emcee_fit} shows the fitting results of the rotational velocity distribution.
The red and green dashed lines represent the halo and disk components, respectively. The vertical 
cyan dashed line represent the rotational velocity 0.

\acknowledgments
We thank Lia Athanassoula and Victor Debattista for helpful discussions.
This work is supported by National Key R\&D Program of 
China No. 2019YFA0405500.
C.L. thanks the National Natural Science Foundation of China (NSFC) with grant No. 11835057.
Y.W. thanks the National Natural Science Foundation of China (NSFC) with grant No. 11773034.
X.-X. Xue thanks the support of NSFC under grants  No. 11873052 and No. 11890694. 
The LAMOST FELLOWSHIP is supported by Special Funding
 for Advanced Users, budgeted and administrated by Center for Astronomical Mega-Science,
 Chinese Academy of Sciences (CAMS). This work is supported by 
 Cultivation Project for LAMOST Scientific Payoff and Research Achievement of CAMS-CAS.
 
Guoshoujing Telescope (the Large Sky Area Multi-Object Fiber Spectroscopic Telescope 
LAMOST) is a National Major Scientific Project built by the Chinese Academy of Sciences.
 Funding for the project has been provided by the National Development and Reform 
 Commission. LAMOST is operated and managed by the National Astronomical Observatories, 
 Chinese Academy of Sciences.
This work has made use of data from the European Space Agency (ESA)
mission {\it Gaia} (\url{https://www.cosmos.esa.int/gaia}), processed by
the {\it Gaia} Data Processing and Analysis Consortium (DPAC,
\url{https://www.cosmos.esa.int/web/gaia/dpac/consortium}). Funding
for the DPAC has been provided by national institutions, in particular
the institutions participating in the {\it Gaia} Multilateral Agreement.

\software{astropy \citep{2013A&A...558A..33A},  \\
          Galpy \citep{2015ApJS..216...29B}, \\
          emcee \citep{emcee}
          }

\bibliography{biblib}{}
\bibliographystyle{aasjournal}

\end{document}